\documentclass[twocolumn,aps,superscriptaddress]{revtex4}
\usepackage{natbib}
\usepackage{amsmath}
\usepackage{amssymb}
\usepackage{graphics}
\usepackage{amsbsy}
\usepackage{bm}

\newcommand{\refeq}[1]{(\ref{#1})}

\newcommand{\be}{{\bf e}}
\newcommand{\bF}{{\bf F}}
\newcommand{\bff}{{\bf f}}

\newcommand{\bI}{{\bf I}}

\newcommand{\bk}{{\bf k}}

\newcommand{\br}{{\bf r}}
\newcommand{\bR}{{\bf R}}

\newcommand{\bT}{{\bf T}}

\newcommand{\bv}{{\bf v}}

\newcommand{\bu}{{\bf u}}
\newcommand{\bU}{{\bf U}}
\newcommand{\bw}{{\bf w}}

\newcommand{\bZ}{{\bf Z}}

\newcommand{\bOmega}{\mbox{\boldmath$\Omega$}}

\newcommand{\brho}{\boldsymbol{\rho}}

\newcommand{\bnabla}{\mbox{\boldmath$\nabla$}}
\newcommand{\btimes}{\boldsymbol{\times}}
\newcommand{\bcdot}{\,\mbox{\boldmath{$\cdot$}}\,}

\newcommand{\ext}{{\rm ext}}

\newcommand{\diff}{{\,\mathrm d}}

\newcommand{\im}{{\mathrm i}}

\newcommand{\smallfrac}[2]{{\textstyle\frac{#1}{#2}}}
\newcommand{\half}{{\textstyle\frac{1}{2}}}
\newcommand{\third}{{\textstyle\frac{1}{3}}}

\newcommand{\const}{\mbox{const}}

\newlength{\tightsecu}
\newlength{\tightsecl}
\newlength{\tightsubsecu}
\newlength{\tightsubsecl}
\newlength{\tightsubsubsecu}
\newlength{\tightsubsubsecl}
\newlength{\sectosubsec}
\newlength{\subsectosubsubsec}
\newcommand{\mathsfb}{\mathsf}
\newcommand{\mathsfbi}{\mathsf}

\newcommand{\remark}[1]{{\tt[#1]}}

\newcommand{\wall}{{\rm w}}
\newcommand{\Wall}{{\rm W}}
\newcommand{\up}{{\rm U}}
\newcommand{\low}{{\rm L}}
\newcommand{\incom}{{\rm in}}

\newcommand{\TW}{{\rm TW}}
\newcommand{\transl}{{\rm t}}
\newcommand{\rot}{{\rm r}}
\newcommand{\parabolic}{\rm p}

\newcommand{\as}{{\rm as}}
\newcommand{\cyl}{{\rm cyl}}

\newcommand{\lmax}{{l_{\max}}}

\newcommand{\sumPrim}[2]{{\sum_{#1}^{#2}}{\hspace{-5pt}\vphantom{\sum}}'
\hspace{5pt}}

\newcommand{\ex}{\hat\be_x}
\newcommand{\ey}{\hat\be_y}
\newcommand{\ez}{\hat\be_z}

\newcommand{\lateralDistance}{\rho}
\newcommand{\lateralVector}{\brho}
\newcommand{\LateralDistance}{\varrho}
\newcommand{\LateralVector}{\boldsymbol{\varrho}}

\newcommand{\identityTensor}{\hat\bI}

\newcommand{\bnablaLat}{\bnabla_\parallel}
\newcommand{\nablaLat}{\nabla_{\parallel}}

\newcommand{\e}{{\rm e}}

\newcommand{\totForce}{\boldsymbol{\mathcal{F}}}
\newcommand{\totTorque}{\boldsymbol{\mathcal{T}}}

\newcommand{\externalVelocity}{\bv^{\rm ext}}
\newcommand{\flowAmplitude}{U_p}
\newcommand{\incidentVelocity}[1]{\bv_{#1}^{\incom}}

\newcommand{\resistanceMatrix}{\boldsymbol{\zeta}}
\newcommand{\resistanceMatrixTP}{\resistanceMatrix^{\transl\parabolic}}
\newcommand{\resistanceMatrixRP}{\resistanceMatrix^{\rot\parabolic}}

\newcommand{\mobilityMatrixP}{\boldsymbol{\nu}}
\newcommand{\mobilityMatrixTP}{\mobilityMatrixP^{\transl\parabolic}}
\newcommand{\mobilityMatrixRP}{\mobilityMatrixP^{\rot\parabolic}}

\newcommand{\mobilityMatrix}{\boldsymbol{\mu}}
\newcommand{\mobilityMatrixTT}{\mobilityMatrix^{\transl\transl}}
\newcommand{\mobilityMatrixTR}{\mobilityMatrix^{\transl\rot}}
\newcommand{\mobilityMatrixRT}{\mobilityMatrix^{\rot\transl}}
\newcommand{\mobilityMatrixRR}{\mobilityMatrix^{\rot\rot}}

\newcommand{\sphericalBasisP}[1]{\bv^+_{#1}}
\newcommand{\sphericalBasisM}[1]{\bv^-_{#1}}
\newcommand{\sphericalBasisPM}[1]{\bv^\pm_{#1}}

\newcommand{\reciprocalSphericalBasisP}[1]{\bw^+_{#1}}
\newcommand{\reciprocalSphericalBasisM}[1]{\bw^-_{#1}}

\newcommand{\CartesianBasisPM}[1]{\bv^\pm_{#1}}

\newcommand{\ScalarBasis}[1]{\Phi_{#1}}
\newcommand{\ScalarBasisP}[1]{\ScalarBasis{#1}^+}
\newcommand{\ScalarBasisM}[1]{\ScalarBasis{#1}^-}

\newcommand{\HeleShawBasis}[2]{\bv^{\as\,#1}_{#2}}
\newcommand{\HeleShawBasisP}[1]{\HeleShawBasis{+}{#1}}
\newcommand{\HeleShawBasisM}[1]{\HeleShawBasis{-}{#1}}

\newcommand{\wallSphericalBasisAsM}[1]{\bu^{\as}_{#1}}

\newcommand{\GreenFreeElement}{G^0}
\newcommand{\GreenWallElement}{G'}
\newcommand{\GreenWallTotElement}{G}
\newcommand{\GreenWallTotElementAs}{G^\as}
\newcommand{\GrandMobilityElement}{M}

\newcommand{\GreenWall}{{\mathsfb G'}}

\newcommand{\GrandMobility}{{\mathsfb M}}
\newcommand{\GrandFriction}{{\mathsfb F}}
\newcommand{\GrandFrictionElement}{F}

\newcommand{\ZsingleWall}{{\mathsfb Z}_\wall}

\newcommand{\tildeCartesianDisplacement}[1]{{\tilde{\mathsfb S}}_{\rm C}^{#1}}
\newcommand{\ScalarDisplacementElements}[1]{S_\cyl^{#1}}

\newcommand{\TransformationCS}[1]{{\mathsfb T}_{\rm CS}^{#1}}
\newcommand{\tildeTransformationCS}[1]{{\tilde{\mathsfb T}}_{\rm CS}^{#1}}
\newcommand{\TransformationSC}[1]{{\mathsfb T}_{\rm SC}^{#1}}

\newcommand{\TransformationElementSAs}{C}

\newcommand{\Kmatrix}{{\mathsfb K}}
\newcommand{\KmatrixElement}{K}

\newcommand{\matrixElementBPM}[1]{B^{\pm}_{#1}}

\newcommand{\TCS}{{\mathsfbi T}_{\rm CS}}
\newcommand{\TSC}{{\mathsfbi T}_{\rm SC}}

\newcommand{\tildeSwp}[1]{{\tilde{\mathsfbi S}}_{\Wall#1}}

\newcommand{\tildeSpw}[1]{{\tilde{\mathsfbi S}}_{#1\Wall}}

\newcommand{\ZW}{{\mathsfbi Z}_\TW}

\newcommand{\inducedForceMultipole}{{\mathsfb f}}
\newcommand{\externalVelocityCoefficient}{{\mathsfb c}}

\newcommand{\resistanceMatrixTT}{\resistanceMatrix^{\transl\transl}}
\newcommand{\resistanceMatrixTR}{\resistanceMatrix^{\transl\rot}}
\newcommand{\resistanceMatrixRT}{\resistanceMatrix^{\rot\transl}}
\newcommand{\resistanceMatrixRR}{\resistanceMatrix^{\rot\rot}}

\newcommand{\vZero}{v_0}
\newcommand{\gammaZero}{\gamma_0}

\newcommand{\epsilonL}{\epsilon_\low}
\newcommand{\epsilonU}{\epsilon_\up}

\begin{document}
\date{\today}

\title{Hydrodynamic interactions of spherical particles in Poiseuille
flow between two parallel walls}

\author{S. Bhattacharya}
\affiliation{Department of Mechanical Engineering, Yale University,
P.O. Box 20-8286, New Haven, CT 06520}
\author{J.\ B{\l}awzdziewicz}
\affiliation{Department of Mechanical Engineering, Yale University,
P.O. Box 20-8286, New Haven, CT 06520}
\author{E.\ Wajnryb}
\affiliation{Department of Mechanical Engineering, Yale University,
P.O. Box 20-8286, New Haven, CT 06520}
\affiliation{On leave from IPPT Warsaw, Poland}

\begin{abstract}
We study hydrodynamic interactions of spherical particles in incident
Poiseuille flow in a channel with infinite planar walls.  The
particles are suspended in a Newtonian fluid, and creeping-flow
conditions are assumed. Numerical results, obtained using our highly
accurate Cartesian-representation algorithm [Physica A xxx, {\bf xx},
2005], are presented for a single sphere, two spheres, and arrays of
many spheres. We consider the motion of freely suspended particles as
well as the forces and torques acting on particles adsorbed at a wall.
We find that the pair hydrodynamic interactions in this wall-bounded
system have a complex dependence on the lateral interparticle distance
due to the combined effects of the dissipation in the gap between the
particle surfaces and the backflow associated with the presence of the
walls.  For immobile particle pairs we have examined the crossover
between several far-field asymptotic regimes corresponding to
different relations between the particle separation and the distances
of the particles from the walls.  We have also shown that the
cumulative effect of the far-field flow substantially influences the
force distribution in arrays of immobile spheres.  Therefore, the
far-field contributions must be included in any reliable algorithm for
evaluating many-particle hydrodynamic interactions in the
parallel-wall geometry.

\end{abstract}

\maketitle

\section{Introduction}
\label{Introduction}

In his pioneering work (more than eighty years ago) Fa\'xen
\cite{Faxen:1924} considered motion of a spherical particle suspended
in a fluid confined by two parallel walls. A recent, considerable
interest in particle motion in confined geometries has been stimulated
by development of new experimental techniques
\cite{%
Prieve-Luo-Lanni:1987,%
Walz-Suresh:1995,%
Faucheux-Libchaber:1994,%
Lin-Yu-Rice:2000,%
Palberg-Biehl:2003,%
Crocker-Matteo-Dinsmore-Yodh:1999}
and by emerging applications, such as the microfluidic devices
\cite{Whitesides-Stroock:2001}
and technologies for production of microstructured materials by a
self-assembly process
\cite{Subramanian-Manoharan-Thorne-Pine:1999,%
Seelig-Tang-Yamilov-Cao-Chang:2002%
}.

There have been published a number of fundamental experimental and
numerical studies on particle dynamics in channels with parallel
planar walls for suspensions of Brownian
\cite{%
Carbajal_Tinoco-Cruz_de_Leon-Arauz_Lara:1997,%
Acuna_Campa-Carbajal_Tinoco-Arauz_Lara-Medina_Noyola:1998,%
Pesche-Nagele:2000,%
Pesche-Kollmann-Nagele:2001,%
Lancon-Batrouni-Lobry-Ostrowsky:2001,%
Marcus-Schofield-Rice:1999,%
Santana_Solano-Arauz_Lara:2001,%
Santana_Solano-Arauz_Lara:2002,%
Cui-Diamant-Lin-Rice:2004,%
Cohen-Mason-Weitz:2004%
}
and non-Brownian
\cite{%
Durlofsky-Brady:1989,%
Nott-Brady:1994,%
Morris-Brady:1998,%
Singh-Nott:2000,%
Morris:2001%
}
particles.  Some of these studies focused on quasi-two-dimensional phenomena
\cite{%
Acuna_Campa-Carbajal_Tinoco-Arauz_Lara-Medina_Noyola:1998,%
Pesche-Nagele:2000,%
Pesche-Kollmann-Nagele:2001,%
Marcus-Schofield-Rice:1999,%
Santana_Solano-Arauz_Lara:2001,%
Cui-Diamant-Lin-Rice:2004%
},
and some on bulk properties, such as particle migration in the pressure-driven
\cite{%
Durlofsky-Brady:1989,%
Nott-Brady:1994,%
Morris-Brady:1998,%
Morris:2001%
}
or shear \cite{Singh-Nott:2000} flow.

Quantitative numerical studies of wall-bounded suspensions require
efficient methods for evaluation of multi-particle hydrodynamic
interactions in these systems.  Some interesting numerical results
were obtained with the help of the wall-superposition approximation
\cite{Pesche-Nagele:2000,Pesche-Kollmann-Nagele:2001} or by modeling
the walls as arrays of immobile spheres \cite{%
Durlofsky-Brady:1989,%
Nott-Brady:1994,%
Morris-Brady:1998,%
Morris:2001,%
Singh-Nott:2000%
}.  
These approaches seem sufficient for describing certain qualitative features
of wall-bounded suspensions (e.g., in Stokesian-dynamic simulations of
hydrodynamic particle diffusion) but the accuracy of such approximations is
often unknown.  Moreover, in some cases, they entirely miss certain important
phenomena. For example, the superposition approximation reproduces neither
the large transverse resistance coefficient of rigid arrays of spheres
\cite{Bhattacharya-Blawzdziewicz-Wajnryb:2005a} nor the enhanced relative
transverse particle motion, observed by Cui {\it et al.\/}
\cite{Cui-Diamant-Lin-Rice:2004} and independently predicted by our recent
analysis
\cite{
Bhattacharya-Blawzdziewicz-Wajnryb:2005%
}.

To overcome these difficulties, we have developed an accurate
Cartesian-representation method for evaluation of multiparticle hydrodynamic
interactions in wall-bounded suspensions of spheres
\cite{Bhattacharya-Blawzdziewicz-Wajnryb:2005a}.  (A related approach was
also independently proposed by Jones \cite{Jones:2004}.) Our method relies on
expanding the flow in a wall-bounded system using two basis sets of Stokes
flows.  The spherical basis set of multipolar flows is used to describe the
interaction of the fluid with the particles, and the Cartesian basis set is
used to account for the presence of the walls.

In our previous studies, the Cartesian-representation method was
applied to determine the resistance functions for systems of spheres
in quiescent fluid
\cite{%
Bhattacharya-Blawzdziewicz-Wajnryb:2005a,%
Bhattacharya-Blawzdziewicz-Wajnryb:2005,%
Bhattacharya-Blawzdziewicz-Wajnryb:2005b%
}.
In the present paper we extend these results to suspensions in a
pressure-driven external flow.  We note that the one-particle motion in such a
system was investigated by Jones \cite{Jones:2004} and Staben {\it et al.\/}
\cite{Staben-Zinchenko-Davis:2003} but, to our knowledge, no accurate
multiparticle results have been reported so far.

This paper is organized as follows.  In Sec.\ \ref{Particles in parabolic
flow} the system is defined, and in Sec.\ \ref{Cartesian-representation
method} the Cartesian-representation method is summarized.  Our numerical
results for single-particle, two-particle, and multiparticle systems are
described in Sec.\ \ref{Results and discussions}.  Conclusions are presented
in Sec.~\ref{Conclusions}.

\section{Particles in parabolic flow}
\label{Particles in parabolic flow}

We consider a suspension of $N$ spherical particles of diameter $d=2a$
in creeping flow between two parallel planar walls.  The no-slip
boundary conditions are satisfied on the walls and the particle
surfaces.  The walls are in the planes $z=0$ and $z=H$, where $H$
denotes the wall separation, and $(x,y,z)$ are the Cartesian
coordinates.  The position of the center of particle $i$ (where
$i=1,\ldots,N$) is denoted by $\bR_i$, and its translational and
rotational velocities are $\bU_i$ and $\bOmega_i$, respectively.  The
external forces and torques acting on particle $i$ are denoted by
$\totForce_i$ and $\totTorque_i$.

In this paper we focus on particle motion in an imposed parabolic flow
of the form
\begin{equation}
\label{parabolic flow}
\externalVelocity
   =4\flowAmplitude \frac{z}{H}\left(1-\frac{z}{H}\right)\ex,
\end{equation}
where $\flowAmplitude$ is the flow amplitude, and $\ex$ is the unit
vector along the $x$ coordinate.  The forces and torques on immobile
particles with 
\begin{equation}
\label{fixed particles}
\bU_i=0,\qquad\bOmega_i=0,
\end{equation}
can be represented by the resistance formula
\begin{equation}
\label{resistance formula for fixed particles}
\left[
   \begin{array}{c}
      \totForce_i\\
      \totTorque_i
   \end{array}
\right]
=-\left[
   \begin{array}{c}
      \resistanceMatrixTP_i\\
      \resistanceMatrixRP_i\\
   \end{array}
\right] \flowAmplitude.
\end{equation}
Similarly, the velocities of freely suspended particles with
\begin{equation}
\label{freely suspended particles}
\totForce_i=0,\qquad\totTorque_i=0,
\end{equation}
can be represented by the mobility formula
\begin{equation}
\label{mobility formula for freely suspended particles}
\left[
   \begin{array}{c}
      \bU_i\\
      \bOmega_i
   \end{array}
\right]
=\left[
   \begin{array}{c}
      \mobilityMatrixTP_i\\
      \mobilityMatrixRP_i\\
   \end{array}
\right] \flowAmplitude.
\end{equation}
In our considerations we assume that the applied flow \refeq{parabolic flow}
is in the $x$ direction.  Thus, the resistance coefficients
$\resistanceMatrixTP_i$ and $\resistanceMatrixRP_i$ and the mobility
coefficients $\mobilityMatrixTP_i$ and $\mobilityMatrixRP_i$ are vectors.  For
the external parabolic flow applied in an arbitrary lateral direction, the
corresponding resistance and mobility coefficients have a tensorial character.

Condition \refeq{freely suspended particles} can be obtained by
applying to immobile particles the forces and torques opposite to
those given by equation \refeq{resistance formula for fixed
particles}.  Thus, the resistance and mobility coefficients
$\resistanceMatrix$ and $\mobilityMatrixP$ satisfy the following
relation
\begin{equation}
\label{relation between parabolic-flow resistance and mobility}
\left[
   \begin{array}{c}
      \mobilityMatrixTP_i\\
      \mobilityMatrixRP_i\\
   \end{array}
\right] 
=
\sum_{j=1}^N
\left[
   \begin{array}{cc}
      \mobilityMatrixTT_{ij}&\mobilityMatrixTR_{ij}\\
      \mobilityMatrixRT_{ij}&\mobilityMatrixRR_{ij}\\
   \end{array}
\right]
\bcdot\left[
   \begin{array}{c}
      \resistanceMatrixTP_j\\
      \resistanceMatrixRP_j\\
   \end{array}
\right],
\end{equation}
where $\mobilityMatrix^{AB}_{ij}$
($A,B=\transl,\rot$) are the translational and rotational
components of the usual mobility matrix \cite{Kim-Karrila:1991} for a
system of particles in quiescent fluid between the walls.  The
many-particle translational--rotational mobility matrix
$\mobilityMatrix^{AB}_{ij}$ is the inverse of the
corresponding multi-particle resistance matrix
$\resistanceMatrix^{AB}_{ij}$, i.e.,
\begin{equation}
\label{mobility and resistance are inverse to each other}
\sum_{j=1}^N
\left[
   \begin{array}{cc}
      \mobilityMatrixTT_{ij}&\mobilityMatrixTR_{ij}\\
      \mobilityMatrixRT_{ij}&\mobilityMatrixRR_{ij}\\
   \end{array}
\right]\bcdot
\left[
   \begin{array}{cc}
      \resistanceMatrixTT_{jk}&\resistanceMatrixTR_{jk}\\
      \resistanceMatrixRT_{jk}&\resistanceMatrixRR_{jk}\\
   \end{array}
\right]
=
\left[
   \begin{array}{cc}
      \identityTensor&0\\
      0&\identityTensor\\
   \end{array}
\right],
\end{equation}
where $\identityTensor$ is the identity tensor. 

In  our recent publications 
\cite{Bhattacharya-Blawzdziewicz-Wajnryb:2005a,
Bhattacharya-Blawzdziewicz-Wajnryb:2005,
Bhattacharya-Blawzdziewicz-Wajnryb:2005b}
we have introduced a formalism that allows us to efficiently evaluate
the translational--rotational mobility matrix $\mobilityMatrix$ for a
system of spherical particles confined between two parallel walls.  In
the present paper our method is used to evaluate the friction and
mobility matrices $\resistanceMatrix^{A\parabolic}_i$ and
$\mobilityMatrixP^{A\parabolic}_i$ ($A=\transl,\rot$) associated with
the Poiseuille flow between the walls.

\section{Cartesian and Hele--Shaw representation methods}
\label{Cartesian-representation method}

In this section we summarize the key elements of our
Cartesian-representation method for evaluating the hydrodynamic
friction and mobility matrices in a suspension confined between two
parallel walls.  We also outline our asymptotic results, which rely on
expansion of the far field flow into a Hele--Shaw basis.  The
asymptotic results apply for sufficiently large interparticle
separations.

A detailed description of our technique is presented in 
\cite{Bhattacharya-Blawzdziewicz-Wajnryb:2005a} and in
\cite{Bhattacharya-Blawzdziewicz-Wajnryb:2005b}.  The Hele--Shaw basis
and its relation to the spherical basis
\cite{Cichocki-Felderhof-Schmitz:1988} used in our analysis
\cite{Bhattacharya-Blawzdziewicz-Wajnryb:2005a} is summarized in
Appendices \ref{Appendix on Hele--Shaw basis} and \ref{Transformation
between the Hele--Shaw and spherical basis sets}.

\subsection{Induced-force formulation}
\label{Induced-force formulation}

In our approach, the effect of the suspended particles on the surrounding
fluid is represented in terms of the induced-force distributions
on the particle surfaces
\begin{equation}
\label{induced forces}
\bF_i(\br)=a^{-2}\delta(r_i-a)\bff_i(\br),
\end{equation}
where 
\begin{equation}
\label{definition of r_i}
\br_i=\br-\bR_i
\end{equation}
and $r_i=|\br_i|$.  By definition of the induced force, the flow field
\begin{equation}
\label{flow field produced by induced forces}
\bv(\br)=\externalVelocity(\br)+\sum_{i=1}^N
  \int\bT(\br,\br')\bcdot\bF_i(\br')\diff\br'
\end{equation}
is identical to the velocity field in the presence of the particles
\cite[][]{Cox-Brenner:1967,Mazur-Bedeaux:1974,Felderhof:1976b}.  Here
\begin{equation}
\label{Green's function}
\bT(\br,\br')=\bT_0(\br-\br')+\bT'(\br,\br')
\end{equation}
is the Green's function for the Stokes flow in the wall-bounded
system, $\bT_0(\br)$ is the Oseen tensor (free-space Green's
function), and $\bT'(\br,\br')$ describes the flow reflected from the
walls.

For a system of particles moving with the translational and angular
velocities $\bU_i$ and $\bOmega_i$ in the external flow
$\externalVelocity$, the induced-force distribution \refeq{induced
forces} can be obtained from the boundary-integral equation of the
form
\begin{eqnarray}
\label{boundary-integral equation for induced-force density}
\lefteqn{[\bZ_i^{-1}\bF_i](\br)
   +\sum_{j=1}^N\int
      [(1-\delta_{ij})\bT_0(\br-\br')\phantom{,}}\nonumber\\
&&\rule{0pt}{0pt}+\bT'(\br,\br')]
      \bcdot\bF_j(\br')\diff\br'
   =\bv_i^{\rm rb}(\br)-\externalVelocity(\br),\nonumber\\
&&\rule{150pt}{0pt}\br\in S_i,
\end{eqnarray}
where the rigid-body velocity field 
\begin{equation}
\label{rigid-body velocity of drop i}
\bv_i^{\rm rb}(\br)=\bU_i+\bOmega_i\btimes\br_i
\end{equation}
and the external flow field $\externalVelocity(\br)$ are evaluated on
the surface $S_i$ of particle $i$.  In the boundary-integral equation
\refeq{boundary-integral equation for induced-force density}, $\bZ_i$
denotes the one-particle scattering operator, which is defined by the
relation
\begin{equation}
\label{definition of operator Z}
\bF_i=-\bZ_i(\incidentVelocity{i}-\bv_i^{\rm rb}),
\end{equation}
where $\incidentVelocity{i}$ is the velocity incident to particle $i$.
For specific particle models, explicit expressions for the operator
$\bZ_i$ are known
\cite[][]{%
Jones-Schmitz:1988,%
Cichocki-Felderhof-Schmitz:1988,%
Blawzdziewicz-Wajnryb-Loewenberg:1999%
}.  

The force and torque acting on particle $i$ can be evaluated from the
induced-force distribution using the integrals
\begin{equation}
\label{force and torque}
   \totForce_i=\int\bF_i(\br)\diff\br,
\qquad
   \totTorque_i=\int\br_i\btimes\bF_i(\br)\diff\br.
\end{equation}
The friction matrix \refeq{resistance formula for fixed particles} can
be computed by solving the boundary equation \refeq{boundary-integral
equation for induced-force density} with the external flow
$\externalVelocity$ in the form \refeq{parabolic flow} and no
rigid-body motion, $\bv_i^{\rm rb}=0$.  Similarly, the
translational--rotational friction matrix is obtained by solving
\refeq{boundary-integral equation for induced-force density} with a
nonzero rigid-body motion \refeq{rigid-body velocity of drop i} and no
external flow, $\externalVelocity=0$.

\subsection{Multipolar expansion}
\label{Multipolar expansion}

In our approach, the boundary-integral equation
\refeq{boundary-integral equation for induced-force density} is solved
after transforming it into a linear matrix equation.  The
transformation is achieved by projecting \refeq{boundary-integral equation
for induced-force density} onto a spherical basis of Stokes flows.  We
use here the multipolar representation introduced by Cichocki {\it et
al.\/} \cite{Cichocki-Felderhof-Schmitz:1988}, but we apply a
different normalization to emphasize full symmetry of the
problem
\cite{%
Bhattacharya-Blawzdziewicz-Wajnryb:2005a,%
Bhattacharya-Blawzdziewicz-Wajnryb:2005%
}.

Accordingly, the induced-force distributions at the surfaces of
particles $i=1\ldots N$ are expanded using the basis set of multipolar
force distributions $\reciprocalSphericalBasisP{lm\sigma}(\br_i)$.
Similarly, the flows incoming to each particle are expanded into the
nonsingular basis set of Stokes flows
$\sphericalBasisP{lm\sigma}(\br_i)$.  Here $l$ and $m$ are the angular
and azimuthal spherical-harmonics orders, and $\sigma=0,1,2$
characterizes the type of the flow.  Explicit definitions of the basis
sets $\reciprocalSphericalBasisP{lm\sigma}$ and
$\sphericalBasisP{lm\sigma}$ (as well as their counterparts
$\reciprocalSphericalBasisM{lm\sigma}$ and
$\sphericalBasisM{lm\sigma}$ that correspond to singular Stokes flows)
are given in
\cite{%
Cichocki-Felderhof-Schmitz:1988,%
Bhattacharya-Blawzdziewicz-Wajnryb:2005a}.

In order to obtain the multipolar representation of the
boundary-integral equation \refeq{boundary-integral equation for
induced-force density}, we apply the multipolar expansion
\begin{equation}
\label{induced force in terms of multipoles}
\bF_i(\br)
   =\sum_{lm\sigma}
      f_i(lm\sigma)
            a^{-2}\delta(r_i-a)
         \reciprocalSphericalBasisP{lm\sigma}(\br_i)
\end{equation}
to the induced-force density \refeq{induced forces}.  The external
flow relative to the particle motion is similarly expanded,
\begin{equation}
\label{expansion of external flow}
\bv_i^{\rm rb}(\br)-\externalVelocity(\br)
   =\sum_{lm\sigma}c_i(lm\sigma)\sphericalBasisP{lm\sigma}(\br_i).
\end{equation}
Inserting these expansions into Eq.\ \refeq{boundary-integral equation
for induced-force density} yields a linear equation of the form
\begin{equation}
\label{induced force equations}
   \sum_{j=1}^N\sum_{l'm'\sigma'}
      \GrandMobilityElement_{ij}(lm\sigma\mid l'm'\sigma')
      f_j(l'm'\sigma')
      =c_i(lm\sigma),
\end{equation}
where the matrix $\GrandMobilityElement$  can be decomposed as
\begin{widetext}
\begin{equation}
\label{Grand Mobility matrix}
      \GrandMobilityElement_{ij}(lm\sigma\mid l'm'\sigma')
   =
      \delta_{ij}\delta_{ll'}\delta_{mm'}Z_i^{-1}(l;\sigma\mid\sigma')
   +
      (1-\delta_{ij})\GreenFreeElement_{ij}(lm\sigma\mid l'm'\sigma')
   +
      \GreenWallElement_{ij}(lm\sigma\mid l'm'\sigma').
\end{equation}
\end{widetext}
The first term on the right side of the above expression corresponds
to the single-particle scattering operator $\bZ_i^{-1}$ in equation
\refeq{boundary-integral equation for induced-force density}; the
second one to the integral operator with the kernel $\bT_0$, and the
third one to the integral operator with the kernel $\bT'$.  Explicit
expressions for the first two terms were derived by Felderhof and his
collaborators
\cite{Jones-Schmitz:1988,%
Cichocki-Felderhof-Schmitz:1988,%
Felderhof-Jones:1989%
} 
some time ago.  Quadrature relations 
\cite{%
Bhattacharya-Blawzdziewicz-Wajnryb:2005a,%
Bhattacharya-Blawzdziewicz-Wajnryb:2005%
}
and asymptotic formulas 
\cite{%
Bhattacharya-Blawzdziewicz-Wajnryb:2005b%
}
for the wall contribution $\GreenWallElement_{ij}$ were recently
derived by our group (as discussed in Sec.\ \ref{Cartesian
representation} below).

\subsection{Friction and mobility of spheres in parabolic flow}
\label{Friction and mobility of spheres in parabolic flow}  
 
In order to evaluate the resistance tensors $\resistanceMatrixTP_i$
and $\resistanceMatrixRP_i$ for immobile particles in Poiseuille flow,
Eq.\ \refeq{induced force equations} is solved with the right-hand
side corresponding to the velocity field \refeq{parabolic flow}.  The
resulting induced-force multipolar distributions \refeq{induced force
in terms of multipoles} are projected onto the total force and torque
using expressions \refeq{force and torque}.  The solution can be
conveniently expressed in terms of the grand friction matrix
\begin{equation}
\label{grand mobility and grand friction}
\GrandFriction=\GrandMobility^{-1},
\end{equation}
which is inverse to the grand mobility matrix $\GrandMobility$ with
the elements given by Eq.\ \refeq{Grand Mobility matrix}.  

As shown in \cite{Bhattacharya-Blawzdziewicz-Wajnryb:2005a}, the
translational--rotational friction matrix
$\resistanceMatrix^{AB}_{ij}$ ($A,B=\transl,\rot$) is given by the relation
\begin{equation}
\label{usual physical friction and grand friction}
\resistanceMatrix^{AB}_{ij}
  =\sum_{lm\sigma}\sum_{l'm'\sigma'}{\bf X}(A\mid lm\sigma)
   \GrandFrictionElement_{ij}(lm\sigma\mid l'm'\sigma')
   {\bf X}(l'm'\sigma'\mid B).
\end{equation}
Here $\GrandFrictionElement_{ij}(lm\sigma\mid l'm'\sigma')$ are the
elements of the grand friction matrix \refeq{grand mobility and grand
friction}, and ${\bf X}(A\mid lm\sigma)={\bf X}^{*}(lm\sigma\mid A)$
are the elements of projection matrices onto the subspace of
translational ($l=1$, $\sigma=0$) and rotational ($l=1$, $\sigma=1$)
rigid-body motions. Explicit
expressions for these matrices are listed in Appendix B of Ref.\
\onlinecite{Bhattacharya-Blawzdziewicz-Wajnryb:2005a}.

The resistance coefficients $\resistanceMatrix^{A\parabolic}$
($A=\transl,\rot$) are given by a relation analogous to \refeq{usual
physical friction and grand friction},
\begin{widetext}
\begin{equation}
\label{friction coeff and grand friction}
\resistanceMatrix^{Ap}_{i}=\sum_{j=1}^N\sum_{lm\sigma}\sum_{l'm'\sigma'}
  {\bf X}(A\mid lm\sigma)\GrandFrictionElement_{ij}(lm\sigma\mid l'm'\sigma')
    Y_{j}(l'm'\sigma'\mid p),
\end{equation}
\end{widetext}
where $Y_{j}(l'm'\sigma'\mid p)$ are the elements of the matrix
representing the orthogonal projection onto the subspace of
pressure-driven flows \refeq{parabolic flow}.  Relation
\refeq{friction coeff and grand friction} and explicit expressions for
the matrix $Y_{j}(l'm'\sigma'\mid p)$ are derive in Appendix
\ref{Projection matrix Y}.  We note that, unlike Eq.\ \refeq{usual
physical friction and grand friction}, relation \refeq{friction coeff
and grand friction} involves summation over the particles.  This
summation is needed because the external parabolic flow
\refeq{parabolic flow} is applied to all particles in the system.

\subsection{Cartesian representation}
\label{Cartesian representation}

To determine the resistance coefficients \refeq{usual physical
friction and grand friction} and \refeq{friction coeff and grand
friction}, the matrix \refeq{Grand Mobility matrix} in the
force-multipole equation \refeq{induced force equations} has to be
first evaluated.  Explicit expressions for the single-particle
scattering matrix $Z_i^{-1}$ and the free-space contribution
$\GreenFreeElement_{ij}$ are known
\cite{%
Cichocki-Felderhof-Schmitz:1988,%
Felderhof-Jones:1989%
}.
To evaluate the wall contribution $\GreenWallElement_{ij}$ to the
matrix \refeq{Grand Mobility matrix} we employ our recently developed
Cartesian-representation method
\cite{Bhattacharya-Blawzdziewicz-Wajnryb:2005a}.  For sufficiently
large interparticle separations appropriate asymptotic expressions
\cite{Bhattacharya-Blawzdziewicz-Wajnryb:2005b} can also be used.

As explained in
\cite{%
Bhattacharya-Blawzdziewicz-Wajnryb:2005a,%
Bhattacharya-Blawzdziewicz-Wajnryb:2005%
}, 
the Cartesian-representation method relies on transformations between
the spherical basis sets of Stokes flows $\sphericalBasisPM{lm\sigma}$
and the Cartesian basis sets $\CartesianBasisPM{\bk\sigma}$ (where {\bk} is
a lateral wave vector).  According to the discussion in Sec.\
\ref{Friction and mobility of spheres in parabolic flow}, the
multipolar spherical sets $\sphericalBasisPM{lm\sigma}$ correspond to
an expansion of the velocity field into spherical harmonics.  Due to
symmetry, the matrix $Z_i$, describing interaction of the flow field
with a spherical particle, is thus diagonal in the spherical-harmonics
orders $l$ and $m$.  The Cartesian basis sets correspond to an
expansion of the velocity field into lateral Fourier modes.  In the
Cartesian representation the matrix $Z_\wall$ that describes
interaction of the flow with a wall is diagonal in the wave vector
$\bk$.  This diagonal structure of the scattering matrices $Z_i$ and
$Z_\wall$ yields a significant simplification of the problem.

To express our results in a compact form, we introduce a matrix
notation in the three-dimensional linear space with the components
corresponding to the indices $\sigma=0,1,2$ that identify the
tensorial character of the basis flow fields
$\sphericalBasisPM{lm\sigma}$.  In this notation, a column vector with
components $a(\sigma)$ is denoted by $\mathsfb a$, and a matrix with
elements $A(\sigma\mid\sigma')$ is denoted by $\mathsfb A$.
Accordingly, the column vectors associated with the coefficients
$f_i(lm\sigma)$ and $c_i(lm\sigma)$ are represented by
$\inducedForceMultipole_i(lm)$ and
$\externalVelocityCoefficient_i(lm)$, and the two-wall Green's matrix
with the elements $\GreenWallElement_{ij}(lm\sigma\mid l'm'\sigma')$
is represented by \mbox{$\GreenWall_{ij}(lm\mid l'm')$}.  We will also
use $3\times6$, $6\times6$ and $6\times3$ matrices composed of
$3\times3$ blocks, as indicated below.

Our result for the wall Green's matrix
$\GreenWallElement_{ij}$ can be expressed in terms of the following
Fourier integral
\begin{equation}
\label{expression for two wall G' -- Fourier integral}
\GreenWall_{ij}(lm\mid l'm')=
   \int\diff\bk\,
   \tilde\Psi(\bk;Z_i,Z_j,H)
   \e^{\bk\bcdot\LateralVector_{ij}},
\end{equation}
where $\LateralVector_{ij}=X_{ij}\ex+Y_{ij}\ey$ is the projection of the
vector $\bR_{ij}=\bR_i-\bR_j$ onto the $x$--$y$ plane, and $\bk=k_x\ex+k_y\ey$
is the corresponding two-dimensional wave vector.  The matrix $\tilde\Psi$ in
the above expression depends on the wall separation $H$ and the vertical
coordinates $Z_{i}$ and $Z_{j}$ of the points $i$ and $j$ (measured
with respect to the position of the lower wall).  This matrix is a product of
several simple matrices,
\begin{eqnarray}
\label{expression for two wall G' -- integrand}
\tilde\Psi(\bk;Z_{i},Z_{j},H)&=&
-\eta^{-1}
   \TSC(lm,\bk)\bcdot\tildeSpw{i}(\bk)
      \bcdot\tilde\ZW(\bk)
\nonumber\\
{}&\bcdot&
      \tildeSwp{j}(\bk)\bcdot\TCS(\bk,l'm'),
\end{eqnarray}
where $\eta$ is the fluid viscosity.

The component matrices
\begin{equation}
\label{two wall transformation matrices}
   \TCS(\bk,lm)=[\TSC(lm,\bk)]^\dagger=
\left[
\begin{array}{c}
      \TransformationCS{+-}(\bk,lm)
   \\\\
      \TransformationCS{--}(\bk,lm)
\end{array}
\right]
\end{equation}
describe the transformations between the spherical ($\rm S$) and
Cartesian ($\rm C$) basis fields.  The superscripts $\pm$ refer to the
singular and nonsingular basis fields for the spherical basis, and the
fields that exponentially grow ($+$) or decay ($-$) in the vertical
direction $z$ for the Cartesian basis.  The matrices \refeq{two wall
transformation matrices} consist of two $3\times3$ blocks
corresponding to the lower and the upper wall, respectively.

Next, the matrices
\begin{equation}
\label{two wall displacement}
   \tildeSwp{s}(\bk)=[\tildeSpw{s}(\bk)]^\dagger=
\left[
   \begin{array}{cc}
      \tildeCartesianDisplacement{++}(kZ_{\low s})&0
\\\\
      0&\tildeCartesianDisplacement{--}(kZ_{\up s})
   \end{array}
\right]
\end{equation}
correspond to the propagation of the Cartesian flow-field components
between the point $s=i,j$ and the lower ($\low$) and upper ($\up$)
walls. Here $Z_{\low s}=-Z_s$ and $Z_{\up s}=H-Z_s$ are the relative
vertical coordinates of the point $s$ with respect to the walls.

\begin{figure}
  \includegraphics{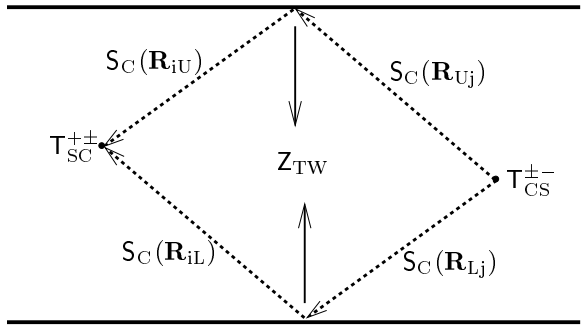}
\caption{
Schematic representation of Eq.\ (\protect\ref{expression for two wall
G' -- integrand}).  The vectors $\bR_{Aj}$ and $\bR_{iA}$ represent
the relative position of the particle $k=i,j$ and the lower ($A=\low$)
or upper ($A=\up)$ wall.
}
\label{schematic}
\end{figure}

Finally, the matrix
\begin{equation}
\label{two wall Z matrix}
   \tilde\ZW(\bk)=
\left[
   \begin{array}{cc}
      \ZsingleWall^{-1}&\tildeCartesianDisplacement{++}(-kH)
\\\\
      \tildeCartesianDisplacement{--}(kH)&\ZsingleWall^{-1}
   \end{array}
\right]^{-1},
\end{equation}
describes scattering of the Cartesian flow components from the walls.
The $3\times3$ matrices $\ZsingleWall$ represent scattering of
the flow from a single wall, and the matrices
$\tildeCartesianDisplacement{++}(-kH)$ and
$\tildeCartesianDisplacement{--}(kH)$ show the propagation of the
flow field between the walls during the multiple-reflection process.
The structure of the expressions \refeq{expression for two wall G' --
Fourier integral}--\refeq{two wall Z matrix} is schematically
represented in Fig.\ \ref{schematic}.  The explicit expressions for
the component matrices \refeq{expression for two wall G' --
integrand}--\refeq{two wall Z matrix} are listed in Appendix
\ref{Expressions for component matrices}.

We note that due to symmetries of the transformation and displacement
matrices and the symmetry
\begin{equation}
\label{symmetry of two wall scattering matrices}
\tilde\ZW(\bk)=[\tilde\ZW(\bk)]^\dagger
\end{equation}
of the two-wall scattering matrix, the Lorentz symmetry of the
two-wall Green's matrix \refeq{expression for two wall G' -- Fourier
integral} is explicit.


\subsection{Far-field asymptotic form}
\label{Far-field form}

The exact Cartesian representation \refeq{expression for two wall G'
  -- Fourier integral}--\refeq{two wall Z matrix} of the wall
contribution to the Green's matrix $\GreenWallElement_{ij}$ involves a
two-dimensional Fourier integral, which has to be evaluated
numerically.  However, for sufficiently large interparticle
separations the calculation can be greatly simplified by using the
far-field asymptotic expressions derived in
\cite{Bhattacharya-Blawzdziewicz-Wajnryb:2005b}.  Below we summarize
this result.

The derivation of the asymptotic expressions for the Green's matrix
\begin{equation}
\label{total Green's function}
\GreenWallTotElement_{ij}=\GreenFreeElement_{ij}+\GreenWallElement_{ij}
\end{equation}
relies on the observation that for large lateral interparticle
distances, $\rho_{12}\gg H$, the disturbance flow scattered from the
particles assumes the Hele--Shaw (i.e., lubrication) form.
Accordingly, the the far-field disturbance flow ${\bf v}^{\rm as}$ is
driven by a two-dimensional harmonic pressure field $p^{\rm as}$,
\begin{equation}
\label{Hele-Shaw flow}
{\bf v}^{\rm as}=-\half\eta^{-1}z(H-z)\bnablaLat p^{\rm as}.
\end{equation}
The pressure $p^{\rm as}$ is independent of the vertical variable $z$
and satisfies the two-dimensional Laplace's equation
\begin{equation}
\label{2D Laplace for Hele-Shaw pressure}
\nablaLat^{2}p^{\rm as}(\lateralVector)=0,
\end{equation}
where $\lateralVector=(x,y)$ is the lateral position with respect to
the particle, and $\bnablaLat$ is the two-dimensional gradient
operator with respect to the lateral coordinates.  The result
\refeq{Hele-Shaw flow} can be obtained using a lubrication expansion
of the Stokes equations in the small parameter $H/\lateralDistance$
(where $\lateralDistance=|\lateralVector|$).

To obtain the asymptotic expression for the Green's matrix
$\GreenWallTotElement_{ij}$ we use the results listed in Appendices
\ref{Appendix on Hele--Shaw basis} and \ref{Transformation between the
Hele--Shaw and spherical basis sets}.  Accordingly, the asymptotic
flow produced by a force multipole \refeq{force multipole} centered at
the position of particle $j$ is expressed in terms of the Hele--Shaw
basis \refeq{Hele-Shaw basis velocity fields} using relation
\refeq{expression for multipolar asymptotic field in Hele-Shaw basis}.
The resulting Hele--Shaw multipolar flow is translated to the position
of particle $i$ using the displacement formula \refeq{displacement
theorem for Hele-Shaw fields}.  Finally the Hele--Shaw field is
transformed back into the spherical basis using relation
\refeq{expansion of Hele-Shaw basis field into spherical basis}.  The
above procedure \cite{Bhattacharya-Blawzdziewicz-Wajnryb:2005b} yields
a compact expression of the form
\begin{widetext}
\begin{equation}
\label{asymptotic Green's matrix elements}
\GreenWallTotElementAs_{ij}(lm\sigma\mid l'm'\sigma')
=-\frac{6}{\pi\eta H^3}
\TransformationElementSAs(Z_i;lm\sigma)
   \ScalarDisplacementElements{+-}
      (\LateralVector_{ij};m\mid m')
         \TransformationElementSAs(Z_j;l'm'\sigma'),
\end{equation}
\end{widetext}
where the component matrices $\TransformationElementSAs$ and
$\ScalarDisplacementElements{+-}$ are given by Eqs.\ \refeq{expression
for scalar displacement matrix} and \refeq{condition for nonzero
elements of C matrix}--\refeq{coefficient A}.  As explained in
\cite{Bhattacharya-Blawzdziewicz-Wajnryb:2005b}, the correction
\begin{equation}
\label{correction to asymptotic result}
\delta\GreenWallTotElement_{ij}
=\GreenWallTotElement_{ij}-\GreenWallTotElementAs_{ij}
\end{equation}
to the asymptotic result \refeq{asymptotic Green's matrix elements}
decays exponentially with the lateral interparticle distance
$\LateralDistance_{ij}$ on the lengthscale $H$.  Typically, the
asymptotic approximation
$\GreenWallTotElement_{ij}\approx\GreenWallTotElementAs_{ij}$ yields
accurate results for $\LateralDistance_{ij}/H\gtrsim2$.


\begin{table*}
\begin{tabular}{l@{\extracolsep{6pt}}cccccccc}
\hline
\hline
\multicolumn{9}{c}{$d/H$}\\
\cline{3-9}
$Z/a$&&0.999&0.995&0.990&0.950&0.900&0.500&0.200\\[3pt]
\hline
1.1&&&&&&0.641&0.583&0.286\\
1.01&&&&0.418&0.498&0.520&0.401&0.188\\
1.007&&&&0.415&0.483&0.502&0.382&0.179\\
1.005&&&0.376&0.409&0.469&0.486&0.366&0.171\\
1.001&&0.304&0.350&0.368&0.409&0.419&0.306&0.141
\end{tabular}
\caption{
Normalized translational velocity $U_{x}/\flowAmplitude$ of a single
sphere of diameter $d=2a$ in imposed parabolic flow
(\protect\ref{parabolic flow}), for different wall separations $H$ and
particle positions $Z$ with respect to the lower wall.  
}
\label{table of linear velocity}
\end{table*}


\begin{table*}
\begin{tabular}{l@{\extracolsep{3pt}}ccccc@{\extracolsep{4pt}}ccc}
\hline
\hline
\multicolumn{9}{c}{$d/H$}\\
\cline{3-9}
$Z/a$&&0.999&0.995&0.990&0.950&0.900&0.500&0.200\\[3pt]
\hline
1.1&&&&&&0.0197&0.723&1.189\\
1.01&&&&5.14E-4&0.101&0.177&0.620&0.903\\
1.007&&&&0.0159&0.109&0.181&0.600&0.866\\
1.005&&&1.95E-4&0.0269&0.115&0.184&0.582&0.834\\
1.001&&2.34E-5&0.0362&0.0556&0.127&0.183&0.504&0.705
\end{tabular}
\caption{
Normalized angular velocity $H\Omega_y/\flowAmplitude$ of a single
sphere of diameter $d=2a$ in imposed parabolic flow
(\protect\ref{parabolic flow}), for different wall separations $H$ and
particle positions $Z$ with respect to the lower wall.  }
\label{table of angular velocity}
\end{table*}


\subsection{Numerical implementation}
\label{Numerical algorithm}  

In order to determine the resistance matrices \refeq{usual physical
friction and grand friction} and \refeq{friction coeff and grand
friction}, the induced-force-multipole equation \refeq{induced force
equations} is solved with the matrix \refeq{Grand Mobility matrix}
evaluated using known results
\cite{Cichocki-Felderhof-Schmitz:1988,Felderhof-Jones:1989} for
$Z_i^{-1}$ and $\GreenFreeElement_{ij}$, and our Cartesian
representation \refeq{expression for two wall G' -- Fourier integral}
for $\GreenWallElement_{ij}$.  For sufficiently large interparticle
distances a simpler relation \refeq{asymptotic Green's matrix
elements} may be be used instead.  After the friction matrices have
been obtained, the mobility matrix \refeq{mobility formula for freely
suspended particles} can be calculated from expressions
\refeq{relation between parabolic-flow resistance and mobility} and
\refeq{mobility and resistance are inverse to each other}.

To accelerate numerical convergence of the Fourier integral in
\refeq{expression for two wall G' -- Fourier integral} (especially,
when both particles $i$ and $j$ are close to a single wall), the
integrand \refeq{expression for two wall G' -- integrand} is
decomposed into two single-wall contributions $\Psi_\low$ and
$\Psi_\up$ and the correction term
\begin{equation}
\label{decomposition of integrand}
\Psi(k)=\Psi_\low(k)+\Psi_\up(k)+\delta\Psi(k).
\end{equation} 
The single wall contributions can be integrated analytically
\cite{Cichocki-Jones-Kutteh-Wajnryb:2000,Cichocki-Jones:1998}.
Moreover, as shown in \cite{Bhattacharya-Blawzdziewicz-Wajnryb:2005b},
the correction term $\delta\Psi(k)$ is easier to integrate numerically
than the original highly oscillatory integrand $\Psi(k)$.

As in other numerical algorithms based on a multipolar expansion of
Stokes flow
\cite{%
Cichocki-Felderhof-Hinsen-Wajnryb-Blawzdziewicz:1994,%
Cichocki-Jones-Kutteh-Wajnryb:2000%
}
the force-multipole equation \refeq{induced force equations} is truncated at a
given multipolar order $l=\lmax$ before it is solved numerically.  To
accelerate the convergence of the approximation with $\lmax$ we employ the
standard lubrication correction \cite{Durlofsky-Brady-Bossis:1987} on the
friction-matrix level.  We closely follow the implementation of the method
described in \cite{Cichocki-Jones-Kutteh-Wajnryb:2000} (for a single wall
problem).  Accordingly, the translational--rotational friction matrix
$\resistanceMatrix_{ij}=\resistanceMatrix^{AB}_{ij}$ ($A,B=\transl,\rot$) is
represented as a combination
\begin{equation}
\label{lubrication superposition for resistance matrix}
\resistanceMatrix_{ij}=
    \resistanceMatrix^{{\rm sup},2}_{ij}
      +\resistanceMatrix^{{\rm sup},\wall}_{ij}
      +\Delta\resistanceMatrix_{ij}
\end{equation}
of the two-particle superposition contribution in free space
$\resistanceMatrix^{{\rm sup},2}_{ij}$, the
single-particle/single-wall superposition contribution
$\resistanceMatrix^{{\rm sup},\wall}_{ij}$, and the remainder
$\Delta\resistanceMatrix_{ij}$.  The superposition contributions
$\resistanceMatrix^{{\rm sup},2}_{ij}$ and $\resistanceMatrix^{{\rm
sup},\wall}_{ij}$ are determined very accurately using the
power-series expansions of the friction matrix in the inverse
interparticle separation and the inverse distance between the particle
and wall, respectively.  The remainder $\Delta\resistanceMatrix_{ij}$,
evaluated as a difference between the multipolar expansion of the full
friction matrix and the superposition contributions, converges with
$\lmax$ much faster than the full friction matrix
$\resistanceMatrix_{ij}$ itself.

In the present implementation of our method, the linear equation
\refeq{Grand Mobility matrix} is solved by matrix inversion.  Thus,
the numerical cost of the calculation scales as $O(N^3)$ with the
number of particles $N$.  (Numerical cost of this order is typical of
unaccelerated Stokesian-dynamics algorithms.)  We note, however, that
the PPPM or fast-multipole acceleration techniques
\cite{Frenkel-Smit:2002} can naturally be used in our
Cartesian-representation algorithm---we will return to this problem in
our future publications.

\section{Results and discussions}
\label{Results and discussions}

We now present some characteristic examples of single- and
many-particle results.  We consider both the motion of freely
suspended particles in the external flow \refeq{parabolic flow} and
forces and torques on fixed particles subjected to this flow.  The
results for an isolated particle are obtained with the truncation at
the multipolar order $l_{max}=32$, which yields accuracy better than
0.1\,\%.  For two-particle and multi-particle systems we use
$l_{max}=12$ and $\lmax=8$, respectively.  The corresponding accuracy
is of the order of 1\,\%.

\subsection{Single particle system}
\label{One particle mobility}

Motion of a single particle in a parabolic flow between two planar
walls was recently considered by Staben {\it et al.\/}
\cite{Staben-Zinchenko-Davis:2003} and by Jones \cite{Jones:2004} (see
also much earlier results by Ganatos {\it et al.\/}
\cite{Ganatos-Pfeffer-Weinbaum:1980}).  We thus give here only limited
results for this system.  In Tables \ref{table of linear velocity} and
\ref{table of angular velocity} we list a set of our highly accurate
results for the linear and angular velocities \refeq{mobility formula
for freely suspended particles} of a force- and torque-free particle
in the parabolic flow \refeq{parabolic flow}.  The linear velocity
$\bU$ is normalized by the magnitude of the parabolic flow
$\flowAmplitude$, and the angular velocity $\bOmega$ by
$\flowAmplitude/H$.  Only the $x$ component of the linear velocity and
the $y$ component of the angular velocity are given because all the
other components vanish by symmetry.

In order to verify our results and test the accuracy of the
calculations reported in \cite{Staben-Zinchenko-Davis:2003}, the
velocities $U^x$ and $\Omega^y$ are given for a subset of
configurations represented in Tables I and II of
\cite{Staben-Zinchenko-Davis:2003}.  We also present some additional
results for tight configurations with $H\approx d$.

We find that our results are in good agreement (up to three digits) with those
reported in \cite{Staben-Zinchenko-Davis:2003} for $Z/a\gtrsim1.01$, where $Z$
is the position of the particle center measured from the lower wall.  For
smaller gaps between the wall and the particle the discrepancies are about
$1.5\%$.  An exception is the rotational velocity in the tightest
configuration reported in \cite{Staben-Zinchenko-Davis:2003} (i.e.,
$d/H=0.95$ and $Z/a=1.007$), where the error is $11\%$.  We expect that these
discrepancies stem from inaccuracies of the boundary-integral calculations in
\cite{Staben-Zinchenko-Davis:2003}---the convergence tests we have performed
indicate that the accuracy of our results is better than $0.05\%$.
We also note that our results agree with those of Jones
\cite{Jones:2004} and with our earlier results of a
multiple-reflection method \cite{Bhattacharya-Blawzdziewicz:2002}.


\begin{figure*}
  \scalebox{0.91}{\includegraphics{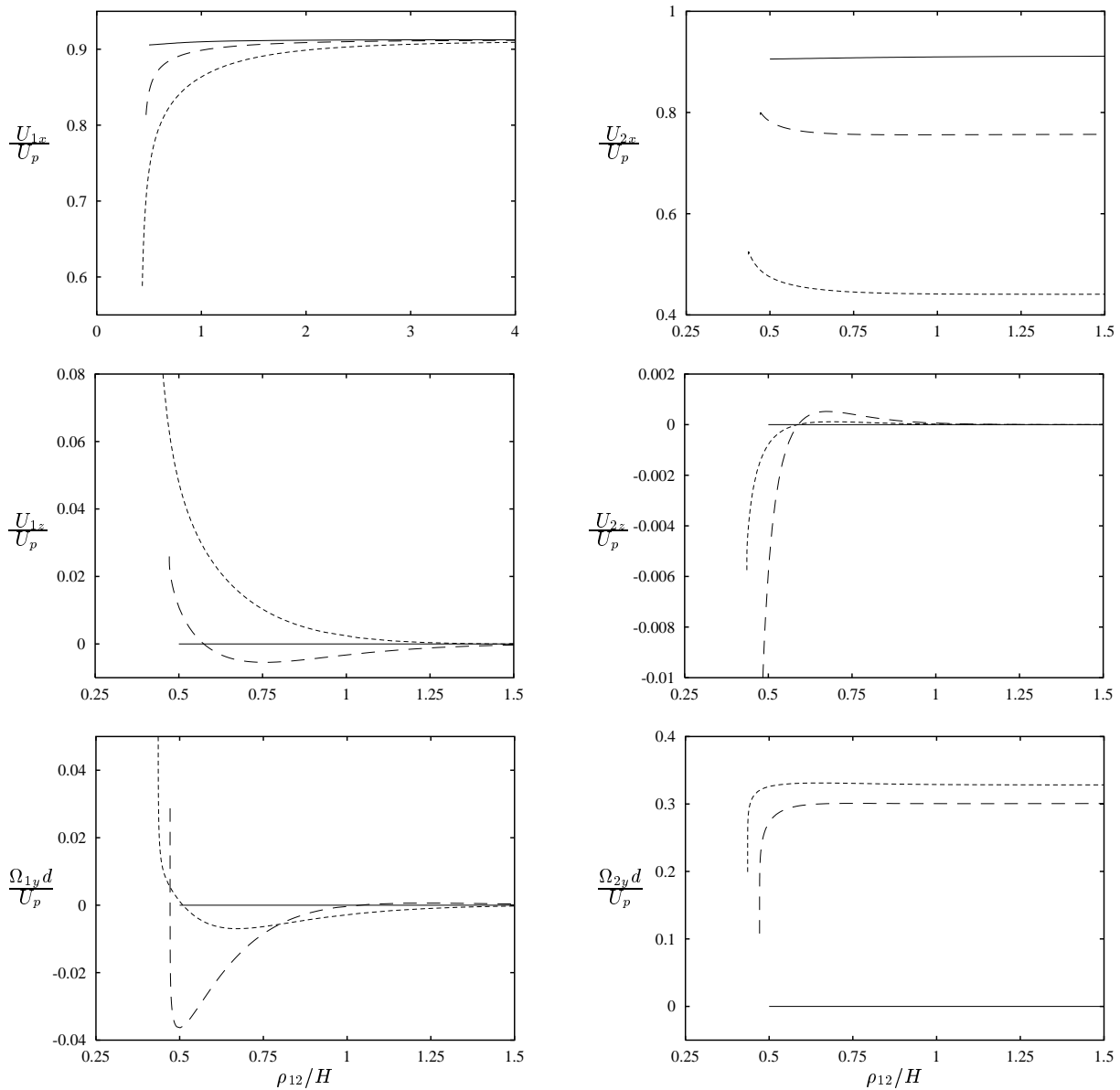}}
\caption{
Normalized linear and angular velocities of two spheres of diameter
$d=2a$ in imposed parabolic flow (\protect\ref{parabolic flow}),
versus particle lateral distance $\rho_{12}$ normalized by wall
separation $H$.  The particle pair has the longitudinal orientation
with respect to the flow direction, the wall separation is $H=2d$, and
sphere $1$ is in the center position $Z_1=2a$.  The positions of
sphere $2$ are $Z_2=1.02a$ (short-dashed line), $Z_2=1.33a$
(long-dashed line), $Z_2=2.0a$ (solid line).   
}
\label{x-h2.0-z1.0}
\end{figure*}


\begin{figure*}
\scalebox{0.91}{\includegraphics{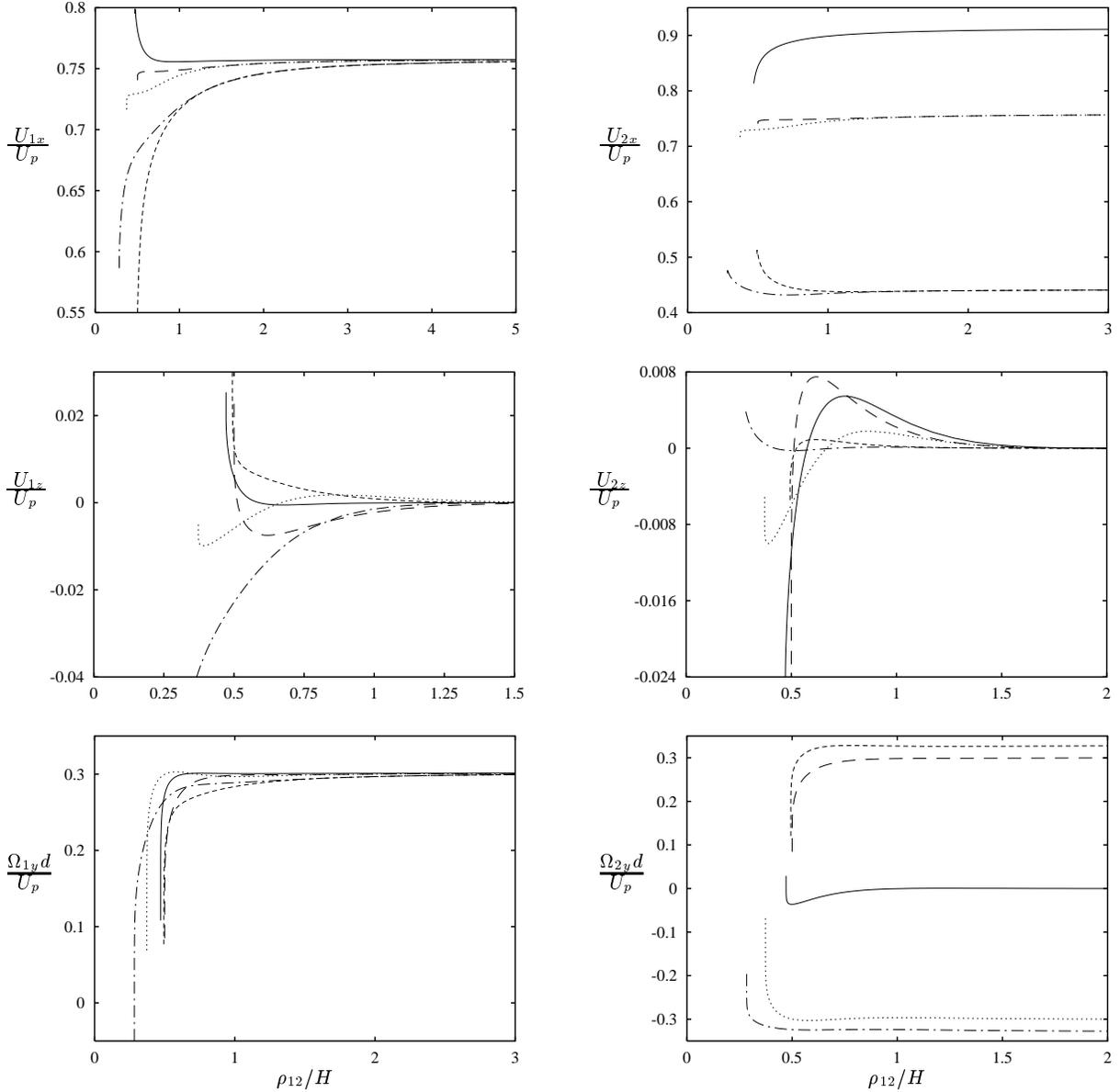}}
\caption{
Same as figure \ref{x-h2.0-z1.0} except that particle $1$ is in an
off-center position $Z_1=1.33a$, and the positions of particle $2$ are
$Z_2=1.02a$ (short-dashed line), $Z_2=1.33a$ (long-dashed line),
$Z_2=2.0a$ (solid line), $Z_2=2.67a$ (dotted line), $Z_2=2.98a$
(dash-dot line).}
\label{x-h2.0-z0.67}
\end{figure*}


\begin{figure*}
  \scalebox{0.91}{\includegraphics{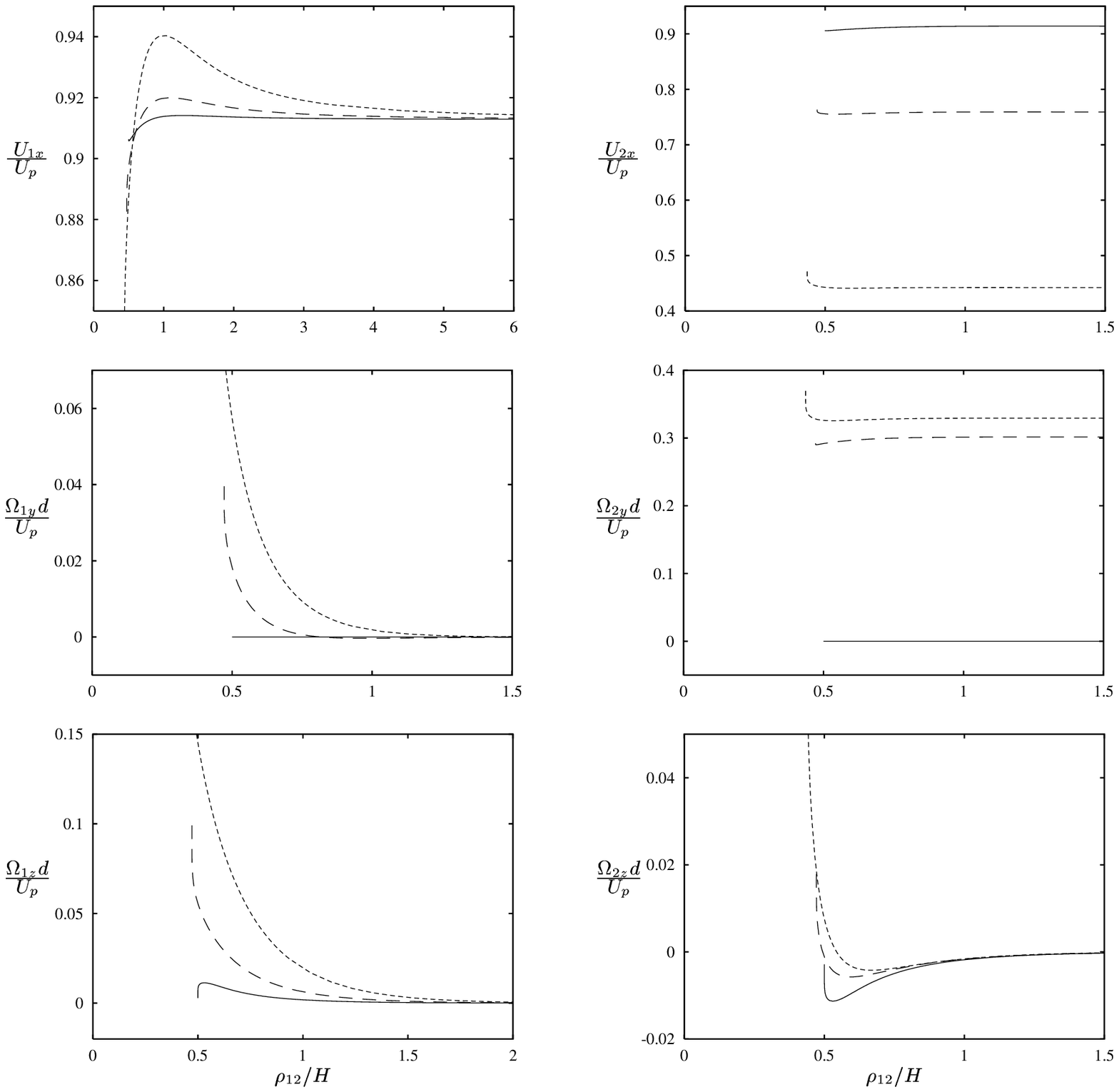}}
\caption{
Same as figure \ref{x-h2.0-z1.0} except that for the transverse
orientation of the particle pair.
}
\label{y-h2.0-z1.0}
\end{figure*}


\begin{figure*}
\scalebox{0.91}{\includegraphics{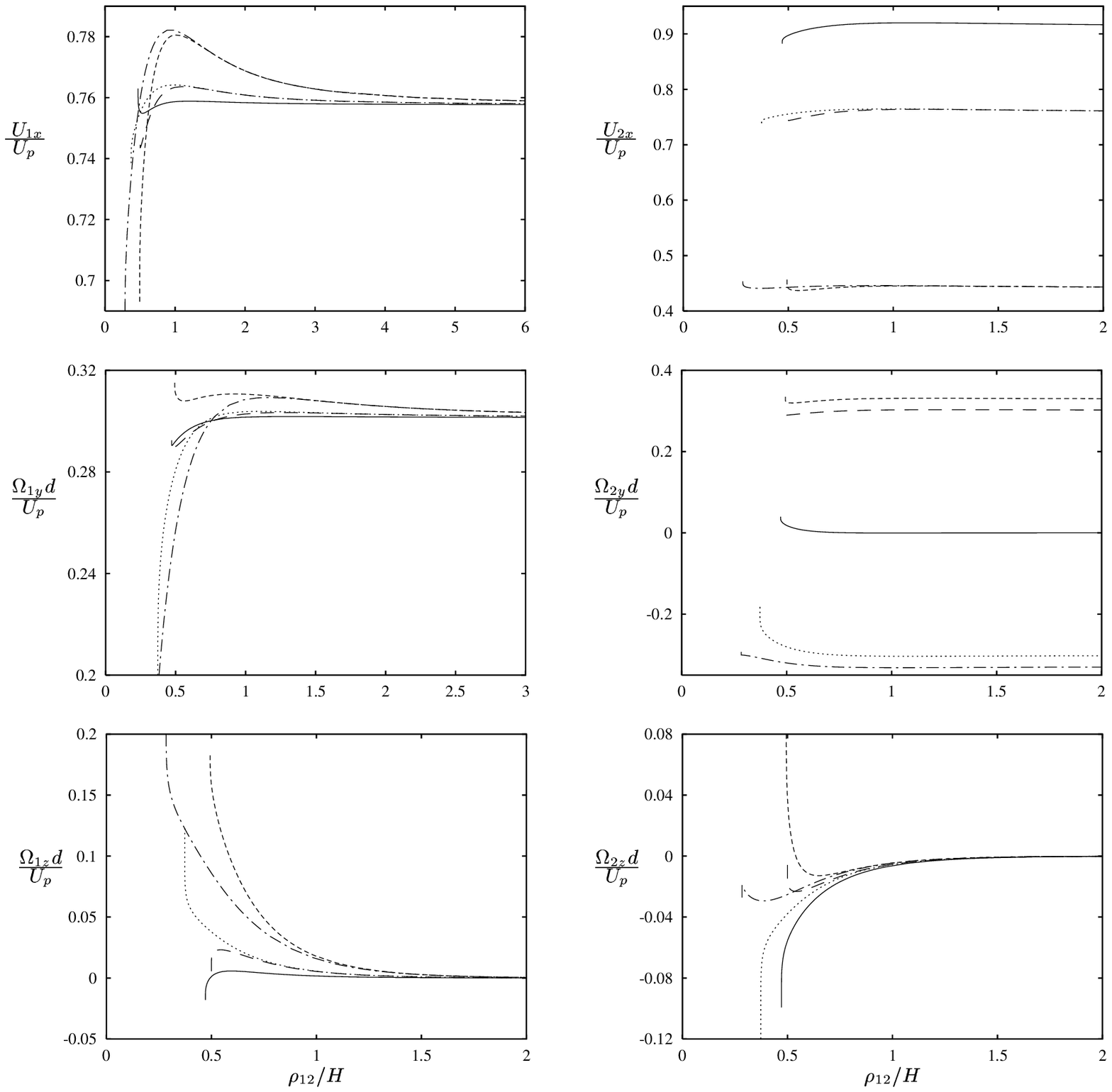}}
\caption{
Same as figure \ref{x-h2.0-z0.67} except that for the transverse
orientation of the particle pair.
}
\label{y-h2.0-z0.67}
\end{figure*}


\subsection{Two-particle system}
\label{Two particles}

\subsubsection{Particle velocities}
\label{Particle velocities}

A sample of characteristic results for the translational and
rotational velocities $\bU_i$ and $\bOmega_i$ (i=1,2) of two force-
and torque-free particles in the parabolic flow \refeq{parabolic flow}
are presented in figures \ref{x-h2.0-z1.0}--\ref{y-h2.0-z0.67}.  The
linear and angular velocities are normalized in these plots by $U_{p}$
and $U_{p}/d$, respectively.  The results are plotted versus the
lateral particle distance $\rho_{12}/H$ for a moderate channel width
\begin{equation}
\label{channel width}
H=2d.
\end{equation}
In Figs.\ \ref{x-h2.0-z1.0} and \ref{y-h2.0-z1.0} the particle $1$ is
in the center position $Z_1=H/2$, and in Fig.\ \ref{x-h2.0-z0.67} and
\ref{y-h2.0-z0.67} it is in the off-center position $Z_1=H/3$.  The
results are given for several vertical positions $Z_2$ of particle
$2$.  (We recall that $Z_i$ denotes the distance of particle $(i)$
from the lower wall.)  In Figs.\ \ref{x-h2.0-z1.0} and
\ref{x-h2.0-z0.67} the particle pair is oriented in the longitudinal
direction $x$ and in Figs.\ \ref{y-h2.0-z1.0} and \ref{y-h2.0-z0.67}
in the transverse direction $y$ with respect to the flow.  We note
that $U_{y}=\Omega_{x}=\Omega_{z}=0$ for the longitudinal
configuration and $U_{y}=U_{z}=\Omega_{x}=0$ for the transverse
configuration, by symmetry.

The results in Figs.\ \ref{x-h2.0-z1.0}--\ref{y-h2.0-z0.67} indicate that the
effect of mutual particle interactions is small if both particles are at the
same vertical position in the channel.  The effect is the largest if one of
the particles is near the channel center and the other close to a wall.  The
results also reveal a different behavior in the near-field and far field
regions, as discussed below.

\paragraph{Near-contact and intermediate region}
\label{Near-contact motion}

The results in Figs.\ \ref{x-h2.0-z1.0}--\ref{y-h2.0-z0.67} indicate
that the dependence of the linear and rotational particle velocities
on the interparticle distance is much more complicated in the
wall-bounded system than in free space.  This complex behavior stems
from the competition between the tangential and normal lubrication
forces and backflow effects associated with the velocity field
scattered from the walls.

For near-contact particle configurations 
\begin{equation}
\label{near-contact configuration}
\epsilon_{12}\ll 1
\end{equation}
(where $\epsilon_{12}=R_{12}/d-1$ is the dimensionless gap between the
particle surfaces, and $R_{12}=|\bR_1-\bR_2|$) the particle dynamics
is strongly influenced by the lubrication forces.  The normal relative
particle motion is arrested by the $O(\epsilon_{12}^{-1})$ normal
lubrication force at the dimensionless gap $\epsilon_{12}$ of several
percent.  The relative tangential and rolling motions are opposed by
much weaker $O(\log\epsilon_{12})$ lubrication forces.  These motions
are thus still quite substantial for $\epsilon_{12}\approx10^{-3}$ and
vanish only for nonphysically small gaps.

A decrease in the relative tangential and rotational particle motion
at small interparticle distances results in an increased overall
dissipation, which may cause a decrease of the horizontal particle
velocities even in symmetric particle configurations with $Z_2=Z_1$ or
$Z_2=H-Z_1$ (cf., the results for $Z_1/H=\third$ and
$Z_2/H=\third,\smallfrac{2}{3}$ in Fig.\ \ref{x-h2.0-z0.67}).  We note
that a pair of touching particles in a transverse configuration
(Figs.\ \ref{y-h2.0-z1.0} and \ref{y-h2.0-z0.67}) does not move, in
general, as a rigid body, because there is no lubrication resistance
to the relative particle rotation around the axis connecting their
centers.

In some cases the normal and tangential lubrication forces have an
opposite effect on a given velocity component.  This produces sharp
kinks in some curves at near-contact positions (e.g., $U_{1z}$ and
$U_{2z}$ for $Z_2=2.67a$ in Fig.\ \ref{x-h2.0-z0.67}).  An additional
change of sign of particle velocities relative to the velocities at
infinite interparticle separations $\rho_{12}\to\infty$ may occur due
to a backflow associated with scattering of the flow from the walls.
Due to a combination of the lubrication and back-flow effects, the $z$
component of the particle velocities changes sign twice for some
longitudinal configurations.

\paragraph{Far-field region}
\label{Intermediate and far field behavior}

As discussed in Sec.\ \ref{Far-field form}, for large lateral
interparticle distances, the hydrodynamic interactions in a
wall-bounded system are determined by the far-field form of the
disturbance flow scattered from the particles.  The scattered flow has
the Hele--Shaw form described by Eqs.\ \refeq{Hele-Shaw flow} and
\refeq{2D Laplace for Hele-Shaw pressure}.  We recall that the
asymptotic form of the flow field is approached exponentially on the
lengthscale $H$.

Taking into account the symmetry of the problem we find that the
far-field disturbance velocity produced by a particle in external flow
\refeq{parabolic flow} is given by equation \refeq{Hele-Shaw flow}
with the pressure of the form
\begin{equation}
\label{solution for Hele-Shaw pressure}
p^{\rm as}\sim \frac{\cos\phi}{\rho},
\end{equation}
where $\phi$ is the polar angle between the lateral position vector
$\brho$ and the flow direction $\ex$.  To the leading order in the
multiple scattering expansion, relations \refeq{Hele-Shaw flow} and
\refeq{solution for Hele-Shaw pressure} determine thus the far-field
form of the hydrodynamic resistance and mobility functions for a pair
of particles in the external parabolic flow.

In particular, Eqs.\ \refeq{Hele-Shaw flow} and \refeq{solution for
Hele-Shaw pressure} indicate that the flow ${\bf v}^{\rm as}$ has only
lateral components.  It follows that the $z$ components of the
translational and rotational particle velocities \refeq{freely
suspended particles} vanish in the far-field regime.  This behavior is
clearly seen in Figs.\ \ref{x-h2.0-z1.0}--\ref{y-h2.0-z0.67}, where
these velocity components approach zero exponentially.

Next, the disturbance field \refeq{Hele-Shaw flow} with the pressure
given by Eq.\ \refeq{solution for Hele-Shaw pressure} behaves as
\begin{equation}
\label{scattered flow decay 2wb}
v^{\rm as}\sim \frac{1}{\rho^{2}}.
\end{equation}
Thus the linear and angular lateral velocities shown in Figs.\
\ref{x-h2.0-z1.0}--\ref{y-h2.0-z0.67} approach the one-particle
asymptotic values as $O(\rho^{-2})$.  The result \refeq{scattered
flow decay 2wb} should be contrasted with the behavior 
\begin{equation}
\label{scattered flow decay infty space}
v^{\rm as}\sim \frac{1}{\rho}
\end{equation}
in free space and 
\begin{equation}
\label{scattered flow decay 1wb}
v^{\rm as}\sim \frac{1}{\rho^{3}}
\end{equation}
in the presence of a single wall (where we assume that
$Z_1,Z_2\ll\rho$).  According to Eqs.\ \refeq{scattered flow decay
2wb}--\refeq{scattered flow decay 1wb}, the decay of the far-field
flow in the presence of one or two walls is faster than the
corresponding decay in free space, because the walls absorb momentum,
and thus they slow the fluid down.

On the other hand, the decay of the flow field \refeq{scattered flow
decay 2wb} in the two-wall system is slower than the decay
\refeq{scattered flow decay 1wb} in the presence of a single wall.
This behavior stems from fluid-volume conservation constraint.  In the
system confined by a single wall the fluid displaced by the particle
primarily flows above the particle, far from the wall, where it
encounters small resistance.  In contrast, in the presence of
two walls, the flow is limited to the quasi-two-dimensional region;
hence, it has a longer range.

Since the total flux associated with the quasi-two-dimensional flow
\refeq{scattered flow decay 2wb} vanishes for $\rho\to\infty$, the
fluid velocity must form a backflow pattern, unlike the behavior in
the unbounded three-dimensional space.  The backflow, described by the
dipolar velocity field \refeq{Hele-Shaw flow} and \refeq{solution for
Hele-Shaw pressure}, results in an enhancement of relative particle
motion for the transverse orientation of the particle pair (as seen
for some particle configurations in the top panels of Figs.\
\ref{y-h2.0-z1.0} and \ref{y-h2.0-z0.67}).  We note that an analogous
behavior was discussed by Cui {\it et al.\/}
\cite{Cui-Diamant-Lin-Rice:2004} in their study of pair diffusion in a
confined, quasi-two-dimensional colloidal suspension.  Similar effect
was also independently described in our recent papers
\cite{%
Bhattacharya-Blawzdziewicz-Wajnryb:2005a,%
Bhattacharya-Blawzdziewicz-Wajnryb:2005%
}.


\begin{figure}
  \includegraphics{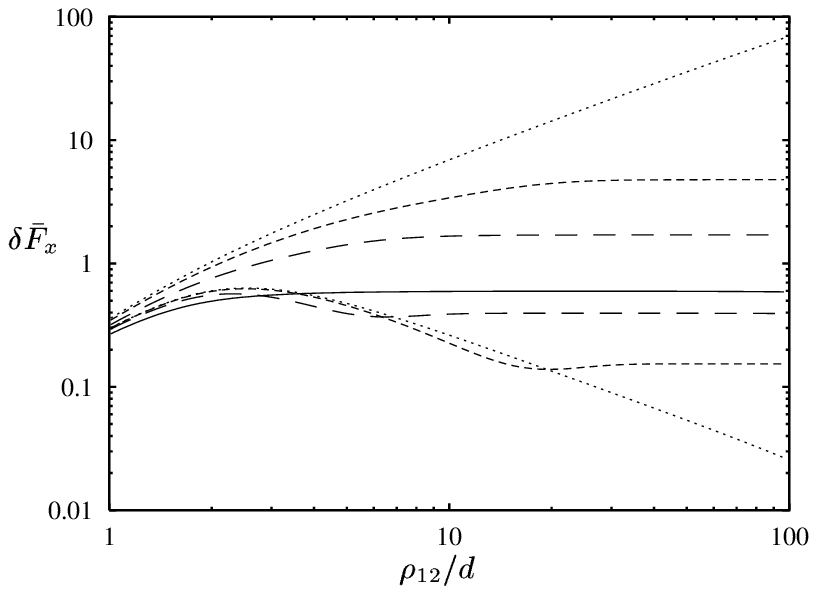}
\caption{
Rescaled force perturbation (\protect\ref{rescaled force
perturbation}) versus interparticle distance $\rho_{12}$ normalized by
the particle diameter $d$ for a pair of particles in longitudinal
orientation.  Wall separation $H=1.02d$ (solid line), $H=5d$
(long-dashed lines), $H=16d$ (short-dashed lines), $H=\infty$ (dotted
lines).  The top three lines correspond to particles in the center
plane $Z_1=Z_2=\half H$, and the bottom three to particles in the
near-wall configuration $Z_1=Z_2=1.02a$.  For $H=1.02d$ (middle line)
the center and near-wall configurations coincide.
}
\label{ff-lng}
\end{figure}


\begin{figure}
  \includegraphics{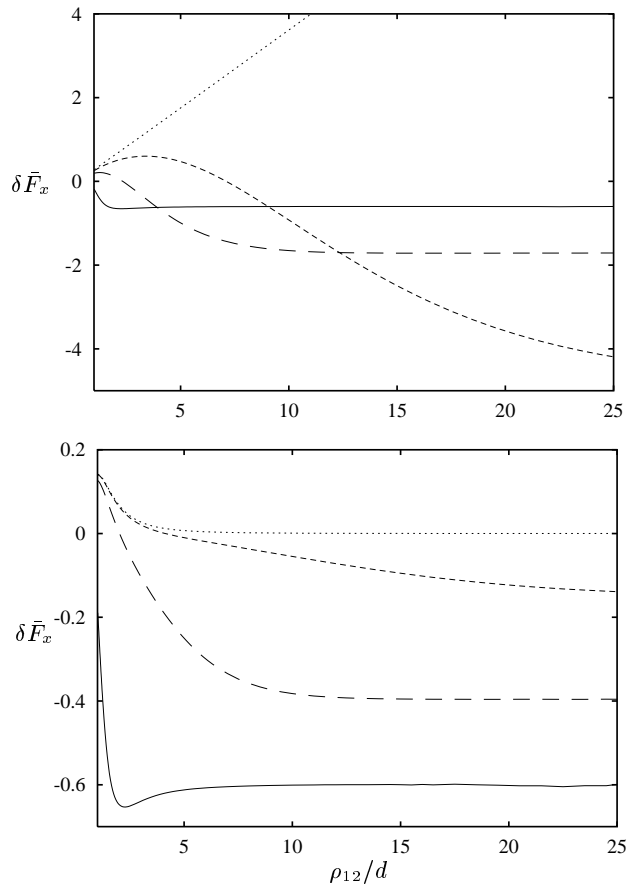}
\caption{
Same as Fig.\ \protect\ref{ff-lng}, except that for the transverse
orientation of the particle pair.  The top panel corresponds to
particles in the center plane $Z_1=Z_2=\half H$, and the bottom panel
to particles in the near-wall configuration $Z_1=Z_2=1.02a$.
}
\label{ff-trn}
\end{figure}


\subsubsection{Crossover behavior}
\label{Crossover behavior}

The far-field disturbance flow, discussed above, affects not only the
velocities of freely suspended particles, but also the hydrodynamic
resistance force \refeq{resistance formula for fixed particles} acting
on immobile particles in the external flow \refeq{parabolic flow}.
Figures \ref{ff-lng} and \ref{ff-trn} illustrate the crossover of the
resistance force between the three regimes corresponding to the
disturbance flows of the form given by equations \refeq{scattered flow
decay 2wb}--\refeq{scattered flow decay 1wb}.  To emphasize the
behavior of the force in the far-field regime, the results are shown
for the $x$ component, $\delta\bar F_{ix}$, of the rescaled force
perturbation
\begin{equation}
\label{rescaled force perturbation}
\delta\bar\bF_i=\left(\frac{\rho_{12}}{d}\right)^2\delta\bF_i,
\end{equation}
where
\begin{equation}
\label{unscaled force perturbation}
\delta \bF_{i}=(\totForce_{i}-\totForce_i^\infty)/F_{\rm st},
\end{equation}
with $F_{\rm st}=3\pi\eta d$ denoting the Stokes resistance force, and
$\totForce_i^\infty$ representing the value of the force
$\totForce_{i}$ for $\rho_{12}\to\infty$.

In Figs.\ \ref{ff-lng} and \ref{ff-trn} force perturbation
$\delta\bar F_{ix}$ is plotted versus the lateral
particle separation $\rho_{12}$ for two particles at the same vertical
position $Z_1=Z_2$.  In one configuration, the particles are at the
center plane
\begin{equation}
\label{two particles in centerplane}
Z_1=Z_2=\half H,
\end{equation}
and in the other one they are close to the lower wall,
\begin{equation}
\label{two particles close to wall}
Z_1=Z_2=1.02a.
\end{equation}
The results are shown for several different wall separations.  Since
the particles are at the same vertical position, the force $\delta\bar
F_x\equiv\delta\bar F_{ix}$ is independent of the particle index $i$.
Figure \ref{ff-lng} represents the results for the longitudinal
orientation of the particle pair, $\brho_{12}=\rho_{12}\ex$, and Fig.\
\ref{ff-trn} the results for the transverse orientation
$\brho_{12}=\rho_{12}\ey$.  The force perturbation \refeq{rescaled
force perturbation} for the longitudinal orientation is positive.  It
is shown on the logarithmic scale to emphasize the algebraic
asymptotic behavior.  For the transverse orientation the perturbation
force in the wall-bounded systems changes sign due to the backflow
effects discussed in Sec.\ \ref{Particle velocities}.  The results are
thus plotted on a linear scale in two separate panels for the center
(top panel) and the near-wall (bottom panel) configurations.

The results shown in Figs.\ \ref{ff-lng} and \ref{ff-trn} clearly
demonstrate the crossover between different regimes corresponding to
the far-field disturbance velocity fields of the form \refeq{scattered
flow decay 2wb}--\refeq{scattered flow decay 1wb}.  For very large
wall separations $H\to\infty$ and the center particle position
\refeq{two particles in centerplane} the rescaled force perturbation
\refeq{rescaled force perturbation} behaves as $\delta\bar
F_x\sim\rho_{12}$, which indicates that $\delta
F_x=O(\rho_{12}^{-1})$, in agreement with the estimate
\refeq{scattered flow decay infty space} of the disturbance-flow
magnitude in free space.  For the near-wall position \refeq{two
particles close to wall} and the longitudinal orientation of the
particle pair we find $\delta F_x=O(\rho_{12}^{-3})$, consistently
with the estimate \refeq{scattered flow decay 1wb}.  In contrast,
$\delta F_x=O(\rho_{12}^{-5})$ in the transverse case, due to an
additional cancellation of the far-field contributions.

For finite wall separations the force perturbation crosses over from
the above-described behavior in the regime $a\ll\rho_{12}\ll H$ to the
behavior $\delta F_x=O(\rho_{12}^{-2})$ (i.e.,
$\delta\bar F_x\sim\const$) for $\rho_{12}\gg H$, in agreement with
Eq.\ \refeq{scattered flow decay 2wb}.  Typically, the far-field
behavior $O(\rho_{12}^{-2})$ is observed already for
$\rho_{12}\gtrsim2H$.

\subsection{Multiparticle systems}
\label{Multiparticle system}

In this section we examine the influence of the walls on the
hydrodynamic interactions in confined multi-particle systems. We focus
on collective phenomena that involve cumulative effects of the
far-field flow \refeq{scattered flow decay 2wb}.  As shown in our
recent studies of particle motion in quiescent fluid
\cite{%
Bhattacharya-Blawzdziewicz-Wajnryb:2005a,%
Bhattacharya-Blawzdziewicz-Wajnryb:2005,%
Bhattacharya-Blawzdziewicz-Wajnryb:2005b%
}, 
the backflow associated with the dipolar form \refeq{Hele-Shaw flow}
and \refeq{solution for Hele-Shaw pressure} of the far-field velocity
may produce a strong, positive feedback resulting in large magnitudes
of induced forces.  In such cases the far-field flow dominates the
behavior of the system.  Below we examine similar phenomena for
particles in the imposed parabolic flow.

\subsubsection{Motion of linear arrays of spheres}
\label{Dynamics of polymer}

First we analyze the effect of confinement on the motion of rigid
linear arrays of touching spheres.  In earlier papers 
\cite{%
Bhattacharya-Blawzdziewicz-Wajnryb:2005a,%
Bhattacharya-Blawzdziewicz-Wajnryb:2005,%
Bhattacharya-Blawzdziewicz-Wajnryb:2005b%
}, 
we have shown that the behavior of such arrays in quiescent fluid is
strongly affected by the walls.  In particular we have demonstrated
that, unlike in free space, the hydrodynamic resistance force in
channels with $H\approx d$ depends significantly on the orientation of
the array with respect to its velocity.  If the orientation of the
array, moving along the channel, is parallel to the velocity, the
resistance force evaluated per one sphere decreases with the length of
the array.  In contrast, for the transverse orientation the resistance
force per particle increases nearly linearly with the array length.
This increase results from the pressure buildup associated with the
positive-feedback backflow effects.


\begin{figure*}
   \scalebox{0.91}{\includegraphics{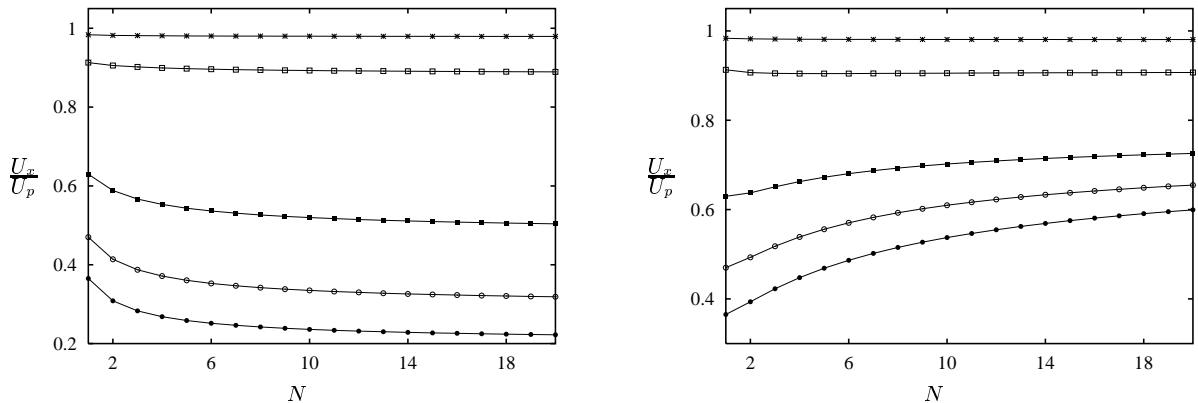}}
\caption{
Normalized translational velocity $U_x/\flowAmplitude$ of rigid,
linear arras of touching spheres, for the longitudinal (left panel)
and transverse (right panel) orientation with respect to the imposed
flow (\protect\ref{parabolic flow}).  The results are shown versus the
number of particles in the array $N$.  The arrays are in the center
plane $z=\half H$ between walls separated by $H=1.004d$ (solid
circles), $H=1.02d$ (open circles), $H=1.1d$ (solid squares), $H=2.0d$
(open squares), $H=4.5d$ (crosses).
}
\label{lng-trn-mid}
\end{figure*}


\begin{figure*}[t]
  \scalebox{0.91}{\includegraphics{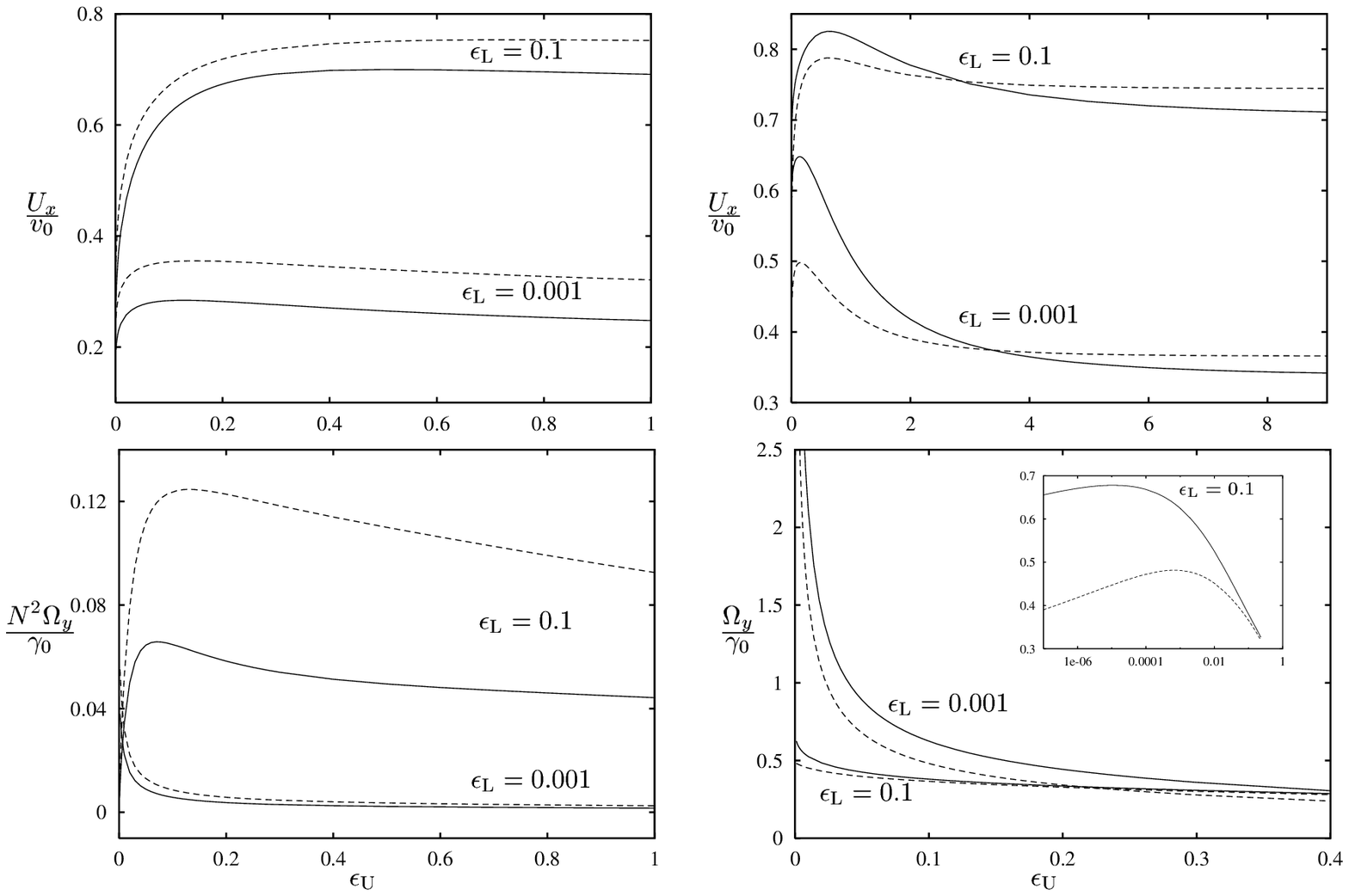}}
\caption{
Normalized translational and angular velocities of rigid, linear arras
of touching spheres with length $N=3$ (dashed lines) and $N=20$ (solid
lines), for the longitudinal (left panel) and transverse (right panel)
orientation with respect to the imposed flow (\protect\ref{parabolic
flow}).  The distance (\protect\ref{normalized distances from the
lower and upper walls}) of the particle surfaces from the lower wall
$\epsilonL$, as labeled; the results are shown versus the distance
from the upper wall $\epsilonU$.  The translational velocity $U_x$ is
normalized by the local fluid velocity \refeq{velocity at center of
particle}; the angular velocity $\Omega_y$ is normalized by the local
shear rate \refeq{shear rate at center of particle}.  For the
longitudinal configuration the angular velocity is in additional
rescaled by $N^2$.  The inset shows the region with small $\epsilonU$.
}
\label{lng-trn-tch}
\end{figure*}


The motion of linear arrays of spheres in the imposed parabolic flow
\refeq{parabolic flow} is illustrated in Fig.\ \ref{lng-trn-mid} and
\ref{lng-trn-tch}.  The arrays are parallel to the walls and are
oriented either in the longitudinal direction $x$ or the transverse
direction $y$.  Figure \ref{lng-trn-mid} presents the translational
velocity of arrays with different length, placed in the mid-plane
$z=\half H$.  The results are given for several channel widths $H$.
Figure \ref{lng-trn-tch} shows linear and angular velocities of arrays
at different vertical positions in the channel.  The linear velocities
are non-dimensionalized by the local velocity of the imposed flow
\begin{equation}
\label{velocity at center of particle}
\vZero=v^\ext(Z)
\end{equation}
and the angular velocities by the local share rate 
\begin{equation}
\label{shear rate at center of particle}
\gammaZero=\frac{\partial v^\ext(Z)}{\partial Z}
\end{equation}
evaluated at the position $Z$ of the array center.  For the mid-plane position
$z=\half H$, $\vZero$ is identical to the amplitude $\flowAmplitude$ of the
imposed flow \refeq{parabolic flow}.

The results in Fig.\ \ref{lng-trn-mid} indicate that the normalized
velocity of an array $U_x/\flowAmplitude$ is smaller in channels with
smaller width.  This behavior stems primarily from the increased
dissipation in the gaps between the particles and the channel walls.
The decrease of the mobility is strongest for long arrays in
longitudinal orientation---the far-field disturbance flow produced by
each of the particles opposes the motion of the array in this case.
For the transverse orientation, the scattered flow acts in the
direction of the external flow; due to the cooperative feedback
effects longer arrays move faster than the shorter ones.  In narrow
channels with $H\approx d$, very long chains in transverse orientation
translate with the velocity that is close to the average velocity of
the unperturbed fluid.

A set of results for short ($N=3$) and a long ($N=20$) linear arrays
at off-center positions in channels with different width is presented
in Fig.\ \ref{lng-trn-tch}.  The configurations are parametrized by
the normalized distances of the particle surfaces from the lower and
upper walls,
\begin{equation}
\label{normalized distances from the lower and upper walls}
\epsilonL=\half(Z-a)/a,\quad\epsilonU=\half(H-Z-a)/a.
\end{equation}
The translational and rotational velocities are shown for arrays at
two vertical positions $\epsilonL=0.001$ and $\epsilonL=1.1$, and they
are plotted versus the distance $\epsilonU$.

The results in the upper panels of Fig.\ \ref{lng-trn-tch} indicate
that the translational velocity of an array at a fixed distance from
the lower wall diminishes rapidly with the decreasing
$\epsilonU\ll\epsilonL$ due to the $O(\log\epsilonU)$ lubrication
resistance associated with the interaction with the upper wall.  In
the case of the longitudinal orientation of the chain, the
translational and rotational velocities saturate at
$\epsilonU\approx1$.  In contrast, for the transverse orientation, the
effect of the upper wall on the translational velocity of the array
has a much longer range, especially for the larger value of the chain
length $N$.  Moreover, the effect of the upper wall is more pronounced
for $\epsilonL=0.001$ than for $\epsilon=0.1$.  These observations are
consistent with the backflow mechanisms discussed above.

Lower panels of Fig.\ \ref{lng-trn-tch} represent the normalized
angular velocity $\Omega_y/\gammaZero$ of the arrays.  We note that
the angular velocity itself changes sign for $\epsilonL=\epsilonU$;
however, the normalized quantities shown in Fig.\ \ref{lng-trn-tch}
are positive, due to the corresponding change of sign of the local
share rate \refeq{shear rate at center of particle}.  For the
longitudinal orientation of the chain the angular velocity is several
orders of magnitude smaller than the angular velocity in the
transverse case.  This strong effect can easily be explained in terms
of particle--wall lubrication forces.  The rotation of the chain
oriented perpendicularly to the flow is governed by the
$O(\log\epsilonL)$ lubrication resistance.  The rotation of the chain
oriented parallel to the flow involves motion of individual particles
towards the wall and away from it, and thus the lubrication forces are
much stronger $O(\epsilonL^{-1})$.  


\begin{figure}
\includegraphics{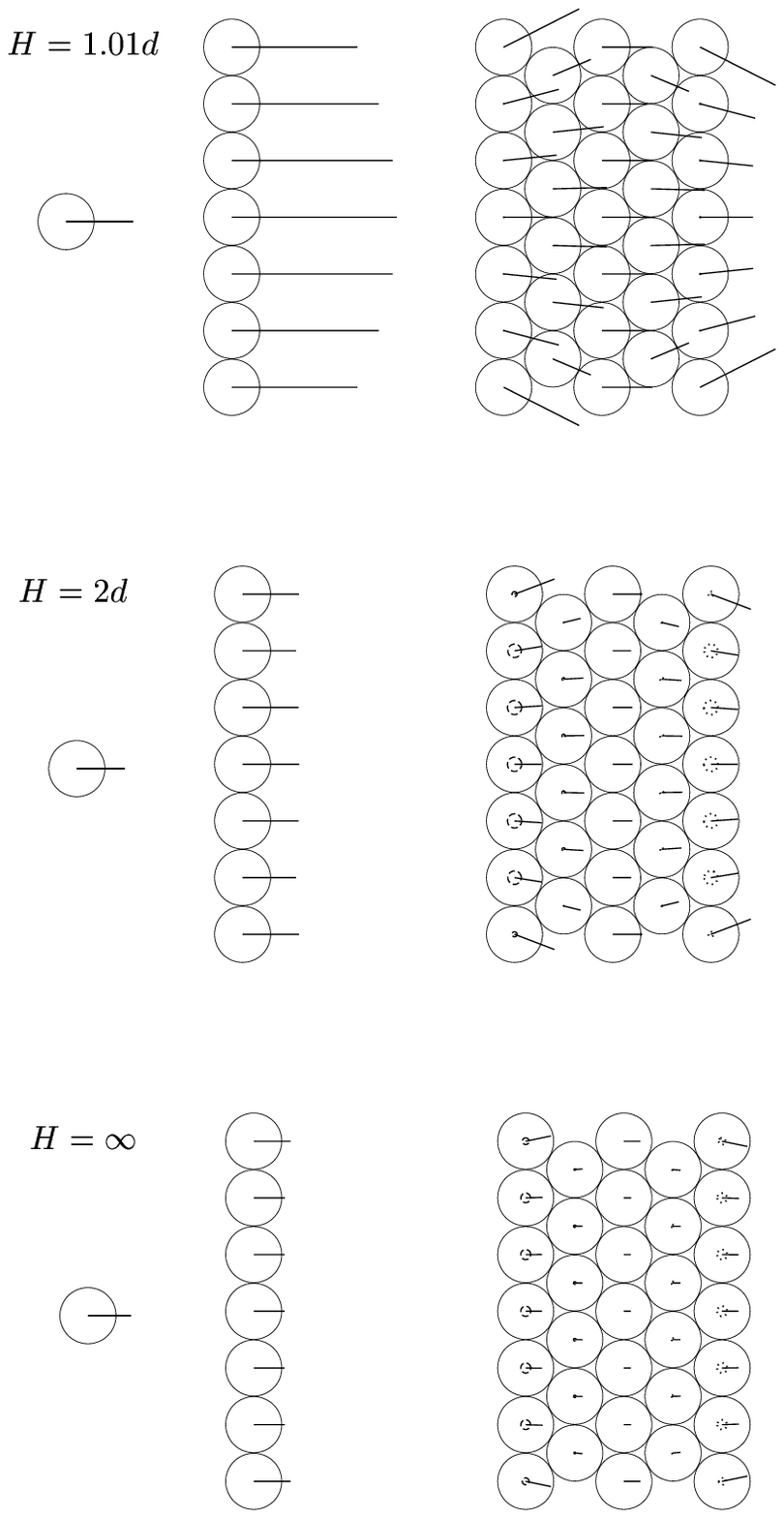}
\caption{
Hydrodynamic drag forces on a single sphere and on individual
particles in linear and hexagonal arrays of spheres adsorbed on a wall
in a channel with the width $H$ (as labeled).  The spheres are
depicted by solid circles.  The lateral forces are represented by the
line segments, and the normal forces by dashed (force away from the
wall) or dotted (towards the wall) circles.  A line segment (circle)
of the length (radius) equal to the particle radius $a$ represents a
force of magnitude $20\pi a\eta\vZero$, where $\vZero$ is the local
fluid velocity (\protect\ref{velocity at center of particle}).
}
\label{grid-force}
\end{figure}


\begin{figure}
\includegraphics{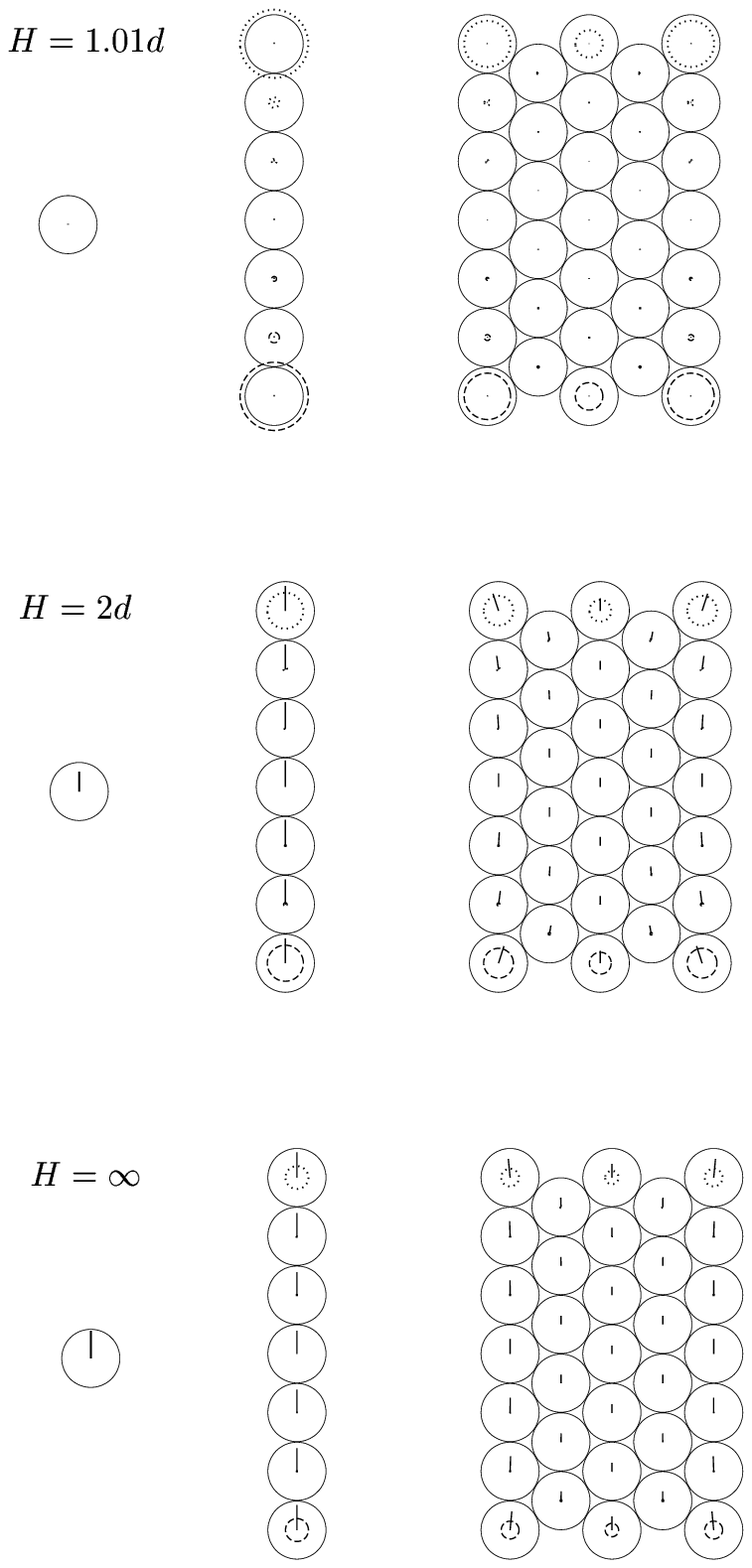}
\caption{
Same as Fig.\ \protect\ref{grid-force}, except that the results are
for the hydrodynamic torque.  A line segment (circle) of the length
(radius) equal to the particle radius $a$ represents a torque of
magnitude $8\pi a^2\eta\vZero$.
}
\label{grid-torque}
\end{figure}


\begin{figure}
\includegraphics{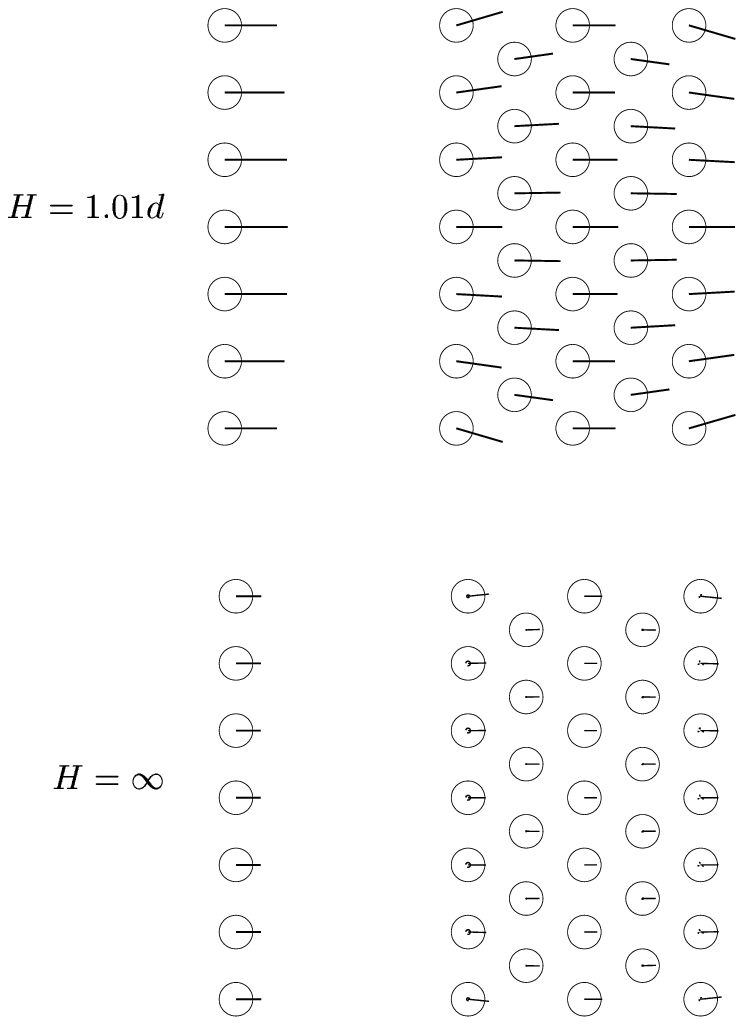}
\caption{
Same as Fig.\ \protect\ref{grid-force} except that the results are
for loosely packed arrays of spheres.
}
\label{grid-force-loose}
\end{figure}


\subsubsection{Hydrodynamic drag on adsorbed particles}
\label{Hydrodynamics of adsorped particles}

In Figs.\ \ref{grid-force}--\ref{grid-force-loose} the results are
presented for the hydrodynamic friction forces and torques on
individual particles in arrays of spheres adsorbed on one of the walls
in a parallel-wall channel.  Understanding of such forces is important
in an analysis of the removal of colloidal particles from a wall by an
applied flow \cite{Ahmadi-Zhang-Han-Greenspan:2005}.  Moreover, our
results provide a further illustration of hydrodynamic phenomena
associated with the far-field form of the disturbance velocity field
produced by the particles.

Figures \ref{grid-force} and \ref{grid-torque} show forces and torques
acting on individual particles in closely packed arrays of touching
spheres.  Arrays that are loosely packed are considered in Fig.\
\ref{grid-force-loose}.  The results are given for a single sphere,
linear arrays of spheres, and hexagonal arrays of spheres.  The
horizontal components of the forces and torques are represented by
line segments and the normal components by the dashed (orientation
away from the wall) and dotted (towards the wall) circles.  The forces
are scaled by $a\eta\vZero$ and the torques by $a^2\eta\vZero$, where
$\vZero$ is the local fluid velocity \refeq{velocity at center of
particle}.  The lengths of the line segments and the radii of the
circles are proportional to the magnitude of the represented quantity.

The results presented in Fig.\ \ref{grid-force} indicate that the
lateral drag force on a single particle only weakly depends on the
channel width, while the forces on particles in linear arrays vary
almost by a factor of five when the wall separation $H$ changes from
$2.02a$ to infinity.  Similarly strong dependence of the forces on the
channel width is observed for the two-dimensional arrays of spheres.

The hydrodynamic drag forces are the largest for linear arrays of
spheres in narrow channels of the width only slightly bigger than the
particle diameter (cf., the top panel of Fig.\ \ref{grid-force}).  The
large forces are associated with the pressure buildup in front of the
array \cite{Bhattacharya-Blawzdziewicz-Wajnryb:2005b}.  As explained
in Sec.\ \ref{Dynamics of polymer}, the pressure buildup involves
interaction of the flow with both walls in an essentially non-additive
manner.  Thus, as shown in
\cite{Bhattacharya-Blawzdziewicz-Wajnryb:2005a}, this effect is not
captured by the usual approximation based on a superposition of two
single-wall contributions.

The results for two-dimensional arrays of spheres involve significant
screening effects resulting from mutual particle interactions.
Accordingly, the lateral forces acting on individual spheres in
two-dimensional arrays are smaller than in the corresponding
linear-array systems.  The forces on the first and last row of
particles are larger than the forces on particles near the center of
the array.  This effect is most pronounced for large wall separations
$H$---in narrow channels the relative force differences are smaller
due to the quasi-two-dimensional character of the flow.

Our results for the vertical forces indicate that their magnitude is
much smaller than the magnitude of the lateral forces. Indeed, the
only significant vertical forces exist on the first and last rows of
particles in two-dimensional arrays.  The maximal value of these
forces occurs for $H\approx4a$.  When the channel width is smaller,
the upper wall suppresses the vertical flow. On the other hand, when
the gap between the top wall and the spheres is too large, the volume
of fluid deflected by the array is distributed over the larger space,
and the vertical flow becomes weaker.  We note that there are no
vertical forces on particles in linear arrays because of the
flow-reversal symmetry.

The behavior of the torque exerted on the particles by the fluid is
illustrated in Fig.\ \ref{grid-torque}.  Characteristic features of
the torque distribution can be explained using arguments similar to
the ones given above.  For example, the lateral torques are the
largest for $H\approx4a$, for the same reason as the corresponding
behavior of the normal force. Our results indicate that the vertical
component of the torque is significant only for particles at the edges
of the arrays, especially for small values of $H$.

Figure \ref{grid-force-loose} shows plots of forces on the spheres in
loosely packed arrays of spheres. For linear arrays, the lateral
forces are relatively small even for $H\approx2a$ because the flow can
pass through the inter-particle gaps without building up a substantial
pressure drop. Moreover, the forces on particles in different
positions in such arrays are of approximately equal magnitude.  For
the two-dimensional loosely packed arrays the lateral forces are
larger than for the closely-packed case, which indicates that the
screening effects are smaller.

\section{Conclusions}
\label{Conclusions}

We have used our recent Cartesian-representation algorithm to study
hydrodynamic interactions of spherical particles in a parabolic flow
between two parallel planar walls.  An important feature of our method
is that at each multipolar-approximation level the boundary conditions
at the walls are exactly satisfied.  This ensures that the far-field
flow produced by the particles has a correct form of a two-dimensional
Hele-Shaw lubrication velocity field.  Our analysis indicates that the
far-filed flow and the associated backflow effects strongly affect
hydrodynamic interactions in confined multiparticle systems.

We have presented a set of numerical results for a single particle, a
pair of particles, and arrays of many particles.  Our one-particle
calculations agree well with earlier results
\cite{%
Staben-Zinchenko-Davis:2003,%
Jones:2004,%
Bhattacharya-Blawzdziewicz:2002}.  
For very tight configurations with small gaps between the particle
surface and the walls we provide more accurate data than those
reported previously \cite{Staben-Zinchenko-Davis:2003}.  For
two-particle and multi-particle systems no accurate results have been
available so far.

Our numerical calculations reveal that the pair and multiparticle
hydrodynamic interactions in the wall bounded system are much more
complex than the interactions in free space.  In particular, unlike in
free space, the sign of mutual friction and mobility functions depends
on the relative particle position in the flow--vorticity plane.  The
changes of sign result form combined effects of the short-range
dissipation in the near-contact regions and backflow due to the
confinement.  Related backflow phenomena were recently observed in
quasi-two-dimensional suspensions of Brownian particles
\cite{Cui-Diamant-Lin-Rice:2004}.  For elongated particles in narrow
slit pores, such a backflow results in a strongly non-isotropic
diffusion constant
\cite{%
Bhattacharya-Blawzdziewicz-Wajnryb:2005a,%
Bhattacharya-Blawzdziewicz-Wajnryb:2005,%
Bhattacharya-Blawzdziewicz-Wajnryb:2005b,%
Han-Alsayed-Nobili-Zhang-Lubensky-Yodh-2006%
}.

The far-field flow also determines the fluid velocity distribution and
the hydrodynamic drag forces in two-dimensional arrays of particles
adsorbed on a wall.  Our results indicate that in narrow channels with
the width $H$ similar to the particle diameter $d$ the hydrodynamic
forces act mostly in the horizontal direction.  Normal forces, which
may lead to particle resuspension, are maximal in channels with
$H\approx2d$.  Our results on hydrodynamic drag on immobile absorbed
particles can be used in an analysis of particle removal from a
channel by a flow.

\begin{acknowledgments}
S.\,B.\ would like to acknowledge the support by NSF grant
CTS-0201131.  E.\,W.\ was supported by NASA grant NAG3-2704 and, in
part, by KBN grant No.\ 5T07C 035 22.  J.\,B.  was supported by NSF
grant CTS-S0348175.
\end{acknowledgments}

\appendix

\section{Hele--Shaw basis}
\label{Appendix on Hele--Shaw basis}

As shown in \cite{Bhattacharya-Blawzdziewicz-Wajnryb:2005b}, the far
filed-flow in the two-wall geometry has the Hele--Shaw, lubrication
form.  Such a flow can be represented in terms of singular ($-$) and
non-singular ($+$) Hele--Shaw basis fields of the form
\begin{equation}
\label{Hele-Shaw basis velocity fields}
\HeleShawBasisP{m}(\br)
   =-\half z(H-z)\bnablaLat\ScalarBasisP{m}(\lateralVector),
\end{equation}
where
\begin{subequations}
\label{Scalar basis}
\begin{equation}
\label{Scalar basis minus}
\ScalarBasisM{0}(\lateralVector)=-\ln\lateralDistance,\qquad
\ScalarBasisM{m}(\lateralVector)=\frac{1}{2|m|}
   \lateralDistance^{-|m|}\e^{\im m\phi}, \quad m\not=0
\end{equation}
are singular and 
\begin{equation}
\label{Scalar basis plus}
\ScalarBasisP{m}(\lateralVector)=
   \lateralDistance^{|m|}\e^{\im m\phi}
\end{equation}
\end{subequations}
non-singular two-dimension harmonic functions.  Here $\lateralVector$
is the lateral position vector with the polar coordinates
$(\lateralDistance,\phi)$, and $\bnablaLat$ is the two-dimensional
gradient operator with respect to the lateral coordinates.  

The Hele--Shaw flow fields \refeq{Hele-Shaw basis velocity fields}
centered at lateral positions $\LateralVector_i$ and
$\LateralVector_j$ are linked by the displacement formula
\begin{equation}
\label{displacement theorem for Hele-Shaw fields}
\HeleShawBasisM{m'}(\br-\LateralVector_j)=
   \sumPrim{m=-\infty}{\infty}
      \HeleShawBasisP{m}(\br-\LateralVector_i)
      \ScalarDisplacementElements{+-}
         (\LateralVector_{ij};m\mid m'),
\end{equation}
where $\LateralVector_{ij}=\LateralVector_i-\LateralVector_j$.  The
term with $m=0$ in the above relation vanishes because
$\HeleShawBasisP{0}\equiv0$ according to Eqs.\ \refeq{Scalar basis
plus} and \refeq{Hele-Shaw basis velocity fields}.  The prime at the
summation sign is introduced to emphasize that this term is omitted.

The matrix 
\begin{equation}
\label{expression for scalar displacement matrix}
\ScalarDisplacementElements{+-}(\LateralVector;m\mid m')
   =\theta(-mm')(-1)^{m'}
\frac{(|m|+|m'|)!}{|m|!|m'|!}
       \ScalarBasisM{m'-m}(\LateralVector)
\end{equation}
in Eq.\ \refeq{displacement theorem for Hele-Shaw fields} is identical
to the displacement matrix for the two-dimensional harmonic potentials
\refeq{Scalar basis}.  We note that due to the presence of the
Heaviside step function
\begin{equation}
\label{Heaviside}
\theta(x)=\left\{
\begin{array}{ll}
0,\qquad&x<0,\\
1,\qquad&x\ge0,
\end{array}\right.
\end{equation}
the Hele--Shaw basis fields with the same sigh of indices $m,m'\not=0$
do not couple in the displacement relation \refeq{displacement theorem
for Hele-Shaw fields}.

\section{Transformation between the Hele--Shaw and spherical basis sets}
\label{Transformation between the Hele--Shaw and spherical basis sets}

In this Appendix we list some relations between the Hele--Shaw basis
of asymptotic far field flows \refeq{Hele-Shaw basis velocity fields}
and the multipolar spherical basis fields defined in
\cite{Cichocki-Felderhof-Schmitz:1988} (in the normalization
introduced in \cite{Bhattacharya-Blawzdziewicz-Wajnryb:2005a}).  

As shown in \cite{Bhattacharya-Blawzdziewicz-Wajnryb:2005a}, the
nonsingular Hele--Shaw field $\HeleShawBasisP{m}$ centered at the
lateral position $\LateralVector_i$ has the following expansion in
terms of non-singular spherical basis fields
$\sphericalBasisP{lm\sigma}$ centered at
$\bR_i=\LateralVector_i+Z_i\ez$,
\begin{equation}
\label{expansion of Hele-Shaw basis field into spherical basis}
\HeleShawBasisP{m}(\br-\LateralVector_i)
   =\sum_{l\sigma}\sphericalBasisP{lm\sigma}(\br-\bR_i)
      \TransformationElementSAs(Z_i;lm\sigma).
\end{equation}
There is also a reciprocal expression 
\begin{equation}
\label{expression for multipolar asymptotic field in Hele-Shaw basis}
\wallSphericalBasisAsM{lm\sigma}(\br-\LateralVector_j;Z_j)
=-\frac{6}{\pi\eta H^{3}}
\HeleShawBasisM{m}(\br-\LateralVector_j)
\TransformationElementSAs(Z_j;lm\sigma)
\end{equation}
for the far-field flow $\wallSphericalBasisAsM{lm\sigma}$ produced,
between the walls, by a force multipole
\begin{equation}
\label{force multipole}
\bF(\br)=a^{-2}\delta(r_j-a)
   \reciprocalSphericalBasisP{lm\sigma}(\br_j)
\end{equation}
(cf., the multipolar expansion \refeq{induced force in terms of
multipoles}).

Explicit expressions for the transformation matrix
$\TransformationElementSAs(Z_i;lm\sigma)$ have been derived in
\cite{Bhattacharya-Blawzdziewicz-Wajnryb:2005b}.  Accordingly, the
nonzero elements of $\TransformationElementSAs(Z_i;lm\sigma)$ satisfy
the condition
\begin{equation}
\label{condition for nonzero elements of C matrix}
l+\sigma-|m|\le2,
\end{equation}  
and they can be written in the form
\begin{equation}
\label{nonzero elements of C matrix}
\TransformationElementSAs(Z;l\,\,\mbox{$\pm\mu$}\,\,\sigma)=
   \matrixElementBPM{l-\mu\,\,\sigma}(Z;\mu),\qquad\mu=|m|\ge1.
\end{equation}
Here, the $\matrixElementBPM{\lambda\,\sigma}(Z;\mu)$
are the elements of the $3\times3$ matrix
\begin{widetext}
\begin{equation}
\label{expression for matrix B}
\left\{
   \matrixElementBPM{\lambda\,\sigma}(Z;\mu)
\right\}_{\lambda,\sigma=0,1,2}
=\half A^{\pm}(\mu)
   \left[
       \begin{array}{ccc}
          -Z(H-Z)&\mp(H-2Z)&2\\
&&\\
          \displaystyle\frac{-\mu(H-2Z)}{(\mu+1)(2\mu+3)^{1/2}}
                    &\displaystyle\frac{\pm2\mu}{(\mu+1)(2\mu+3)^{1/2}}
                   &0\\
&&\\
          \displaystyle\frac{2\mu(\mu+1)^{1/2}}{(\mu+2)(2\mu+3)(\mu+5)^{1/2}}
                   &0&0
       \end{array}
    \right],
\end{equation}
\end{widetext}
with
\begin{equation}
\label{coefficient A}
A^{\pm}(\mu)
  =(\mp 2)^{\mu}\,\mu!\,\left[\frac{4\pi}{(2\mu+1)(2\mu)!}\right]^{1/2}.
\end{equation}
The range $\lambda=0,1,2$ of the index $\lambda=l-|m|$ in equation
\refeq{expression for matrix B} result from the conditions $|m|\le l$
and \refeq{condition for nonzero elements of C matrix}.  

\section{Projection matrix Y}
\label{Projection matrix Y}

Relation \refeq{friction coeff and grand friction} is derived by
inserting expansion \refeq{expansion of external flow} of the external
flow \refeq{parabolic flow} into Eqs.\ (141) and (145) in Ref.\
\cite{Bhattacharya-Blawzdziewicz-Wajnryb:2005a}.  Comparing Eqs.\
\refeq{resistance formula for fixed particles} and \refeq{friction
coeff and grand friction} to the resulting formula we conclude that
the the matrix $Y_{j}(lm\sigma\mid p)$ is determined by the expansion
\begin{equation}
\label{external flow general expansion}
\externalVelocity(\br)=\sum_{lm\sigma}\sphericalBasisP{lm\sigma}({\bf r}-{\bf
R}_{j})Y_j(lm\sigma|p)
\end{equation}
of the parabolic flow \refeq{parabolic flow} into the spherical basis
centered at ${\bf R}_{j}$.  

To obtain the explicit expression for the matrix $Y_{j}(lm\sigma\mid
p)$, we represent the external parabolic flow \refeq{parabolic flow}
in terms of the Hele-Shaw asymptotic basis \refeq{Hele-Shaw basis
velocity fields},
\begin{equation}
\label{external flow and Hele-Shaw basis}
\externalVelocity
   =-4\flowAmplitude H^{-2}(\HeleShawBasisP{-1}+\HeleShawBasisP{1}).
\end{equation}
Inserting expansion \refeq{expansion of Hele-Shaw basis field into
spherical basis} into the above relation and comparing the result to
\refeq{external flow general expansion} yields
\begin{equation}
\label{expression for matrix Y in B}
Y_j(lm\sigma|p)
  =-4\flowAmplitude H^{-2}(\delta_{m1}+\delta_{m\,-1})C(Z_j;lm\sigma),
\end{equation}
where $\delta_{mk}$ denotes the Kronecker delta.  Using relations
\refeq{condition for nonzero elements of C matrix}--\refeq{coefficient
A} for the matrix $C$ we thus find
\begin{widetext}
\begin{equation}
\label{expression for matrix Y}
\left\{Y_j(\lambda+1\ \pm 1\ \sigma|p)\right\}_{\lambda,\sigma=0,1,2}
=-4H^{-2}\flowAmplitude \sqrt{\frac{2\pi}{3}}
   \left[
       \begin{array}{ccc}
          -Z_j(H-Z_j)\,&\mp (H-2Z_j)\,\,&2\\
&&\\
          \displaystyle\frac{-(H-2Z_j)}{2\sqrt{5}}
                    &\displaystyle\frac{\pm1}{\sqrt{5}}
                   &0\\
&&\\
          \displaystyle\frac{2}{15\sqrt{3}}
                   &0&0
       \end{array}
    \right].
\end{equation}
\end{widetext}
All other elements of $Y_j$ vanish, by Eq.\ \refeq{condition for
nonzero elements of C matrix}.  \remark{Sukalyan: verify !?.}

\section{Component matrices in Cartesian representation}
\label{Expressions for component matrices}

In this Appendix we list explicit expressions for the component
matrices in the Cartesian representation \refeq{expression for two
wall G' -- Fourier integral}--\refeq{two wall Z matrix} of the Green's
matrix $\GreenWall_{ij}$.  These expressions are derived in Ref.\
\cite{Bhattacharya-Blawzdziewicz-Wajnryb:2005a}.

The component transformation matrices in Eq.\ \refeq{two wall
transformation matrices} can be represented in the factorized form
\begin{eqnarray}
\label{factorization of transformation CS}
\TransformationCS{\pm-}(\bk,lm)
   =[\TransformationSC{+\pm}(lm,\bk)]^{\dagger}
   =\im^m(2\pi k)^{-1/2}\e^{\im m\psi}&&\nonumber\\
\times
      \tildeTransformationCS{\pm-}(lm)\bcdot\Kmatrix(k,l),&&\hspace{2em}
\end{eqnarray}
where $(k,\psi)$ are the polar coordinates of the vector $\bk$.
In the above expression
\begin{equation}
\label{matrix K}
\KmatrixElement(k,l;\sigma\mid\sigma')=\delta_{\sigma\sigma'}k^{l+\sigma-1},
\end{equation}
and 
\begin{subequations}
\label{form of transformation SC and CS}
\begin{equation}
\label{form of transformation CS}
   \tildeTransformationCS{+-}=
(-1)^{l+m}\left[
\begin{array}{ccc}
      c&-2b&4a\\
      b&-2a&0\\
      a&0&0
\end{array}
\right],
\end{equation}
\begin{equation}
   \tildeTransformationCS{--}=
\left[
\begin{array}{ccc}
      a&0&0\\
      b&2a&0\\
      c&2b&4a
\end{array}
\right].
\end{equation}
\end{subequations}
The three independent scalar coefficients in equations \refeq{form of
transformation SC and CS} are
\begin{subequations}
\label{coefficients a,b,c}
\begin{eqnarray}
\label{coefficient a}
a&=&[4(l-m)!(l+m)!(2l+1)]^{-1/2},\\\nonumber\\
\label{coefficient b}
b&=&2am/l,\\\nonumber\\
\label{coefficient c}
c&=&a\frac{l(2l^2-2l-1)-2m^2(l-2)}{l(2l-1)}.
\end{eqnarray}
\end{subequations}

The component displacement matrices in Eqs.\ \refeq{two wall
displacement} and \refeq{two wall Z matrix} are given by the
relation
\begin{equation}
\label{expression for tilde S}
\tildeCartesianDisplacement{++}(kZ)
   =[\tildeCartesianDisplacement{--}(-kZ)]^\dagger
=\left[
   \begin{array}{ccc}
      1&0&2kZ\\
      0&1&0\\
      0&0&1
   \end{array}
\right]\be^{kZ}.
\end{equation}
For rigid walls with no-slip boundary conditions the single-wall
reflection matrix in Eq.\ \refeq{two wall Z matrix} has the form
\begin{equation}
\label{expression for Z wall}
\ZsingleWall
   =\left[
       \begin{array}{ccc}
          1&0&0\\
          0&1&0\\
          0&0&1
       \end{array}
    \right].
\end{equation}
For planar interfaces with other boundary conditions (e.g., a
surfactant-covered fluid-fluid interface discussed in
\cite{Blawzdziewicz-Cristini-Loewenberg:1999}) the scattering matrix
is different from identity, and it may depend on the magnitude of the
wave vector $k$.  Explicit expressions for scattering matrices for
such systems will be presented elsewhere.


\end{document}